\documentclass[12pt]{article}
\usepackage{hhline}
\usepackage{hyperref}
\usepackage{comment}
\usepackage{color}
\usepackage{amsfonts}
\usepackage{amsmath}
\usepackage{amssymb}
\usepackage{mathtools}
\usepackage{dsfont}
\usepackage{cite}
\usepackage{subcaption}
\usepackage{graphicx}
\usepackage{array}
\usepackage{notoccite}
\usepackage[table]{xcolor}
\usepackage{enumitem}   
\usepackage{multirow}
\usepackage{tabularx}

\usepackage{mathrsfs}
\usepackage{pgfplots}
\pgfplotsset{compat=1.15}
\usepackage{mathrsfs}
\usetikzlibrary{arrows}

\usepackage{fullpage}
\usepackage{geometry}
\renewcommand{\equiv}{:=}

\newcommand{\Dhat}{{\widehat{D}}}
\newcommand{\Nhat}{{\widehat{N}}}
\newcommand{\Ibbhat}{{\widehat{\mathbb{I}}}}
\newcommand{\psihat}{{\widehat{\psi}}}
\newcommand{\kin}{{\rm kin}}

\newcommand{\M}{\text{M} \ }
\newcommand{\LF}{\text{LF} \ }
\newcommand{\SUPP}{\text{SUPP} \ }
\newcommand{\SR}{\text{SR} \ }
\newcommand{\SYM}{\text{SYM} \ }

\title{A minimalist approach to 3D photoemission orbital tomography: algorithms and data requirements
\footnotetext[0]{This research was funded by the Deutsche Forschungsgemeinschaft (DFG, German Research Foundation) – Project-ID 432680300 – SFB 1456, Project B01. }}
\author{
Thi Lan Dinh\thanks{Institut f\"ur Numerische und Angewandte Mathematik, Universit\"at G\"ottingen,%
\ Lotzestr.~16--18, 37083 G\"ottingen, Germany. E-mail: \texttt{t.dinh@math.uni-goettingen.de}}
\and  G. S. Matthijs Jansen\thanks{I. Physical Institute, Universit\"at G\"ottingen, Friedrich-Hund-Platz 1,  37077 G\"ottingen, Germany.
E-mail: \texttt{gsmjansen@uni-goettingen.de}}
\and D. Russell Luke\thanks{Institut f\"ur Numerische und Angewandte Mathematik, Universit\"at G\"ottingen,%
\ Lotzestr.~16--18, 37083 G\"ottingen, Germany. E-mail: \texttt{r.luke@math.uni-goettingen.de}}
\and Wiebke Bennecke\thanks{I. Physical Institute, Universit\"at G\"ottingen, Friedrich-Hund-Platz 1,  37077 G\"ottingen, Germany.
E-mail: \texttt{wiebke.bennecke@stud.uni-goettingen.de}}
\and Stefan Mathias\thanks{I. Physical Institute, Universit\"at G\"ottingen, Friedrich-Hund-Platz 1,  37077 G\"ottingen, Germany.
E-mail: \texttt{smathias@uni-goettingen.de}}}

\date{\today}

\begin{document}

\maketitle
\begin{abstract}
Photoemission orbital tomography provides direct access from laboratory measurements to the real-space molecular orbitals of well-ordered organic semiconductor layers. Specifically, the application of phase retrieval algorithms to photon-energy- and angle-resolved photoemission data enables the direct reconstruction of full 3D molecular orbitals without the need for simulations using density functional theory or the like. A major limitation for the direct approach has been the need for densely-sampled, well-calibrated 3D photoemission patterns. Here, we present an iterative projection algorithm that completely eliminates this challenge: for the benchmark case of the pentacene frontier orbitals, we demonstrate the reconstruction of the full orbital based on a dataset containing only four simulated photoemission momentum measurements. We discuss the algorithm performance, sampling requirements with respect to the photon energy, optimal measurement strategies, and the accuracy of orbital images that can be achieved.
\end{abstract}


\section{Introduction}
Image reconstruction based upon sparse Fourier space measurements is a problem that recurs in a wide range of imaging modalities, with well known examples being X-ray computed tomography \cite{withers_x-ray_2021, piccolomini_reconstruction_2018} and magnetic resonance imaging \cite{lustig_sparse_2007}. In fields such as coherent diffractive imaging \cite{miao_extending_1999, Marchesini_x-ray_2003} and photoemission orbital tomography \cite{puschnig_reconstruction_2009}, the challenge of image reconstruction is further increased due to the inherently incomplete measurement data: whereas image reconstruction requires the complex-valued field to be known, only the amplitude can be measured. For the latter two, iterative image reconstruction methods nowadays enable high-resolution imaging with only limited prior information \cite{marchesini_invited_2007, jansen2020efficient}, but these methods still require strongly oversampled data to achieve this. In many cases, particularly for photoemission orbital tomography which concerns itself with the quantum-mechanical electronic orbitals of molecules, the object of interest is intrinsically sparse, and this can be efficiently exploited using the ideas of compressive sensing surveyed in  \cite{candes_introduction_2008}.

\begin{figure}
    \centering
    \includegraphics[width=0.7\textwidth]{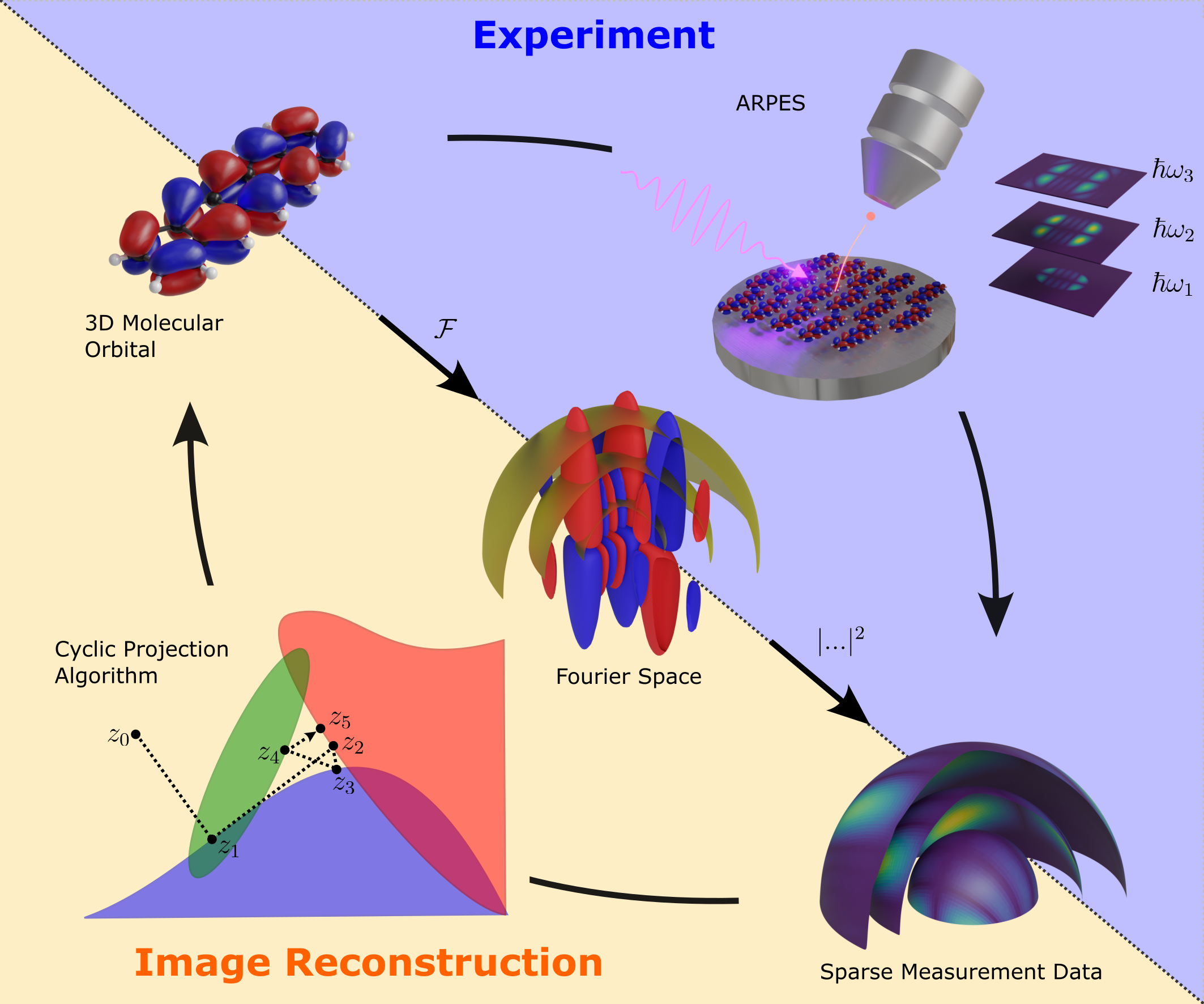}
    \caption{\label{f:schema} Photoemission orbital tomography provides access to the three dimensional structure of molecular orbitals (top left) by combining photon-energy- and angle-resolved photoemission spectroscopy (ARPES) with image reconstruction methods. Here, the photoelectron distribution of an ordered molecular layer measured at different photon energies $\hbar \omega$ (top right) corresponds to the amplitude squared of hemispherical cuts through the three dimensional Fourier space (center). Based on the sparsely sampled Fourier space (bottom right) the initial molecular orbital can be reconstructed using a cyclic projection algorithm (bottom left).}
\end{figure}

Photoemission orbital tomography (POT) is well established as a probe of the electronic structure of thin molecular films. Until recently, the conventional approach of this method to achieve information on the molecular orbitals would be the comparison of angle-resolved photoemission spectroscopy (ARPES) data to simulations generated using density functional theory (DFT) \cite{dauth_orbital_2011, zamborlini_multi-orbital_2017, yang_identifying_2019}.  Iterative phase retrieval, however, allows one to {\em reconstruct} the orbitals directly from the measurements  \cite{puschnig_reconstruction_2009, jansen2020efficient}, and this is independent of DFT or similar calculations (see Fig. \ref{f:schema}).
\if{On the other hand, POT also enables a direct imaging of molecular orbitals through iterative phase retrieval \cite{puschnig_reconstruction_2009, jansen2020efficient}, which is of particular value because it does not depend on DFT or similar calculations. }\fi
While spatially resolved access to the electron density of molecular orbitals can also be achieved, e.g., by scanning tunneling microscopy \cite{Repp_STM_2005, Soe_STM_2009}, the unique nature of POT provides several advantages to probe the molecular orbitals at high resolution. Most notably, together with gas-phase molecular orbital tomography \cite{itatani_tomographic_2004, vozzi_generalized_2011}, POT using photon-energy-dependent ARPES measurements is uniquely suited to image the full three-dimensional (3D) molecular orbitals, and POT remains so far the only technique that can do so for larger, nanometer-scale molecules \cite{graus_three-dimensional_2019, weis_exploring_2015}. Such 3D-POT is highly promising, as the combination with time-resolved photoemission orbital tomography\cite{Wallauer20sci, Neef23nat, baumgartner_ultrafast_2022, Bennecke23arxiv} will access spatio-temporal properties of the dynamic electronic wavefunction that are so far inaccessible by experimental means. Additionally, 3D-POT promises a crucial insight at hybrid interfaces, where strong modifications of the electronic structure are known to occur \cite{yang_momentum-selective_2022, lihuang_organic2d_2018}.

However, despite the rapid development of powerful multi-dimensional photoelectron detection schemes\cite{keunecke_time-resolved_2020, kutnyakhov_time-_2020, heber_multispectral_2022}, their application for 3D-POT has remained limited for a number of reasons: First and foremost, the application of phase retrieval algorithms requires a high-resolution 3D momentum distribution, implying the time-costly measurement of an ARPES dataset with a large number of photon energies, where, furthermore, the photon flux must be accurately calibrated. Moreover, the plane-wave model of photoemission, upon which phase retrieval methods are based, is only an approximation with respect to the overall photon-energy dependence of the photoelectron intensity \cite{kern_simple_2023, kirschner_quantitative_2024}. In this article, we deal with these challenges using a fundamentally new approach for 3D-POT: by implementing ideas from sparse optimization \cite{HesseLukeNeumann14}, we perform an image reconstruction on a pixelized object
that is defined by the combination of several qualitative constraints in addition to sparsity constraints. 

In Section~\ref{sec:theory_POT}, we review the fundamentals of three-dimensional photoemission orbital tomography, and identify the challenges that must be solved to facilitate an efficient access to the 3D molecular orbital. We translate this theoretical description in Section~\ref{sec:theory_reconstruction} to an image reconstruction problem, and explain how this nonconvex problem can be solved by a cyclic projection algorithm with several qualitative constraints. The orbital reconstruction from simulated data for an isolated pentacene molecule is demonstrated in Section~\ref{sec: Numerical results}. After establishing the convergence behavior of cyclic projections in this setting, we focus on two questions: first, how many photon energies must be measured for accurate orbital reconstruction, and second, we investigate how experimental noise such as counting statistics and inaccurately estimated light intensity calibration affect the reconstruction.

\section{Theory of 3D photoemission orbital tomography}
\label{sec:theory_POT}
In image-reconstruction photoemission orbital tomography, we consider an ordered molecular layer in which the unit cell consists of a single molecule (e.g., a monolayer of pentacene on Ag(110)), and use the plane-wave model of photoemission to calculate the ARPES signal. In this case, for a given molecular orbital $\psi$ the ARPES signal $I$ for the photoelectron momentum $\vec{k}$ can be expressed as
\begin{equation}
    I_{\hbar\omega}(\vec{k}, E_{\kin}) = |\vec{A}\cdot \vec{k}|^2~ |\psihat(\vec{k})|^2~\delta(E_{\rm b} + E_{\kin} - \hbar \omega)\,.
    \label{eq:plane_wave_model}
\end{equation}
Here $\vec{A}$ is the vector potential of the ionizing radiation, $\psihat\equiv \mathcal{F}\psi$, the Fourier transform of the molecular orbital, and $|\psihat(\vec{k})|$ denotes the absolute value of $\psihat$ at the point $\vec{k}$ in momentum space. The Dirac $\delta$ function stems from Fermi's golden rule and ensures energy conservation, taking into account the ionization potential $E_{\rm b}$ of the molecular orbital, the final state kinetic energy $E_{\kin} = \hbar^2 k^2/2 m_e$ and the photon energy $\hbar \omega$.

Note that $E_{\kin}$ is proportional to the magnitude squared of the momentum $\vec k$;  hence, a single ARPES measurement provides the amplitude of the orbital's momentum distribution along a hemispherical shell with a radius that depends directly on the photon energy $\hbar \omega$.
In order to probe the full 3D momentum distribution $|\psihat(\vec{k})|$ of the molecular orbital, it is therefore necessary to measure ARPES patterns for multiple photon energies. This naturally leads to a question that is central to the work that we present here: \emph{How many photon energies are required to yield a reliable orbital reconstruction from ARPES momentum maps?} Before it is possible to address this question, however, first a strategy must be established that enables the reconstruction of a 3D molecular orbital from a sufficiently large dataset. 

A reconstruction strategy for 3D photoemission orbital tomography (3D-POT) must solve two problems: 1) Starting with an ARPES dataset that only contains discrete cuts through the 3D photoelectron momentum distribution, the complete 3D photoelectron momentum distribution must be determined, and 2) the phase distribution corresponding to the measured amplitudes must be resolved. Up until now, these steps have been performed sequentially: first, the 3D momentum distribution is completed using an interpolation algorithm \cite{graus_three-dimensional_2019}. Subsequently, the full 3D momentum distribution can then be used as input of an iterative phase retrieval algorithm, which is the standard approach to determine the phase pattern in 2D orbital imaging as well. The interpolation step has significant implications: here, it is (implicitly) assumed that the momentum distribution is of a shape, or {\em regularity},  that can be interpolated and furthermore that it is properly sampled. Such Fourier-space interpolation, however, is prone to errors, particularly when also the intensity-only nature of the measurement is taken in account. Consider for example a zero crossing in the momentum distribution $\psihat$. Here, intensity measurements at opposite sides will both yield non-zero intensity, and interpolation of this data will lead to the false conclusion that the intensity in-between must also be non-zero. The size of the error introduced here decreases as the sampling density increases, meaning that an accurate reconstruction of 3D molecular orbitals through this sequential approach relies on large datasets in which many photon energies have been measured.

In order to reduce the number of ARPES measurements needed and, more importantly, to prevent errors arising from the interpolation procedure, it is thus clear that a completely different reconstruction strategy must be developed. To that end, we now consider the phase retrieval algorithms that are already commonly used in orbital reconstruction photoemission orbital tomography. Generally, these are mathematical optimization algorithms that aim to find a solution that best satisfies a set of constraints. In photoemission orbital tomography, the solution corresponds to the image of a molecular orbital, while the set of constraints consists of the ARPES measurements and the prior knowledge about the molecular orbital. These algorithms are commonly referred to as \emph{phase retrieval algorithms} because they reconstruct the phase that is lost upon the measurement, but the recovered information can include intensity as well as phase. For example, phase retrieval algorithms in coherent diffractive imaging are already routinely used to also recover the intensity in parts of the diffraction pattern that cannot be measured \cite{giewekemeyer_high-dynamic-range_2014, schroer_coherent_2008}. As such, a promising approach to reconstruct 3D molecular orbitals from a discrete set of ARPES measurements is to generalize the already-necessary phase retrieval algorithms to 3D reconstruction algorithms. In this article, we build upon the sparsity-driven phase retrieval algorithm for orbital imaging\cite{jansen2020efficient} to develop a 3D-POT reconstruction algorithm. In particular, using the compact and voxel-sparse shape of the molecular orbital as a real-space constraint, we present an algorithm that efficiently and accurately interpolates the momentum distribution based upon ARPES measurements at just a few photon energies.

\section{Basic approach}
\label{sec:theory_reconstruction}
\subsection{From data to phase retrieval problem}

\begin{figure}[ht] 
\begin{subfigure}{.25\textwidth}
  \centering
  \includegraphics[width=0.8\linewidth]{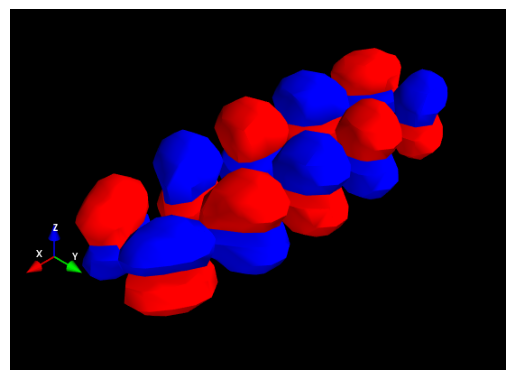}
  \caption{}
  \label{fig:3D.molecule}
\end{subfigure}
\begin{subfigure}{.65\textwidth}
  \centering
  \includegraphics[width=1\linewidth]{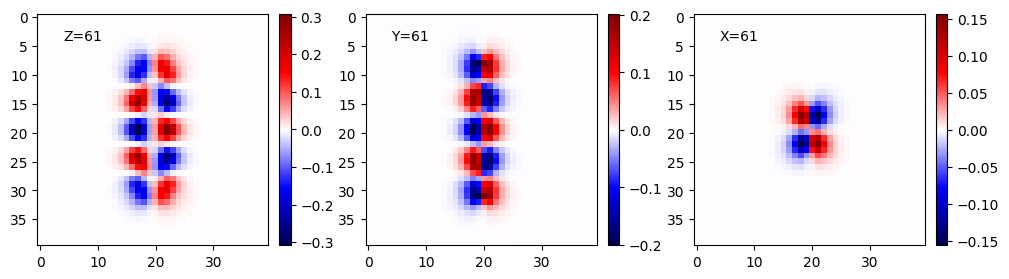}
  \caption{}
  \label{fig:slice.molecule}
\end{subfigure}
\begin{subfigure}{.25\textwidth}
  \centering
  \includegraphics[width=.8\linewidth]{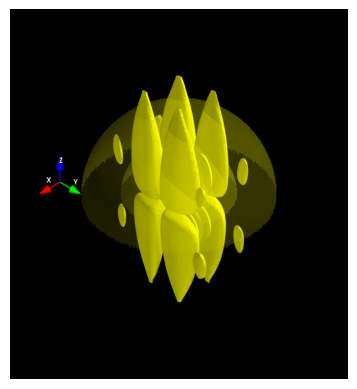}
  \caption{}
  \label{fig:abs.FT.molecule}
\end{subfigure}
\begin{subfigure}{.65\textwidth}
  \centering
  \includegraphics[width=1.0\linewidth]{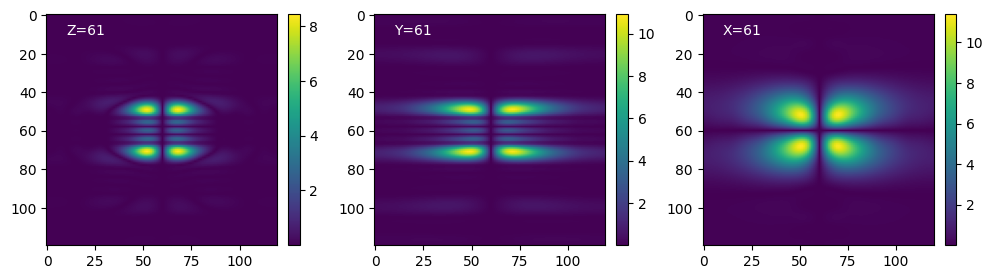}
  \caption{}
  \label{fig:slices.abs.FT.molecule}
\end{subfigure}
\begin{subfigure}{.25\textwidth}
  \centering
  \includegraphics[width=.8\linewidth]{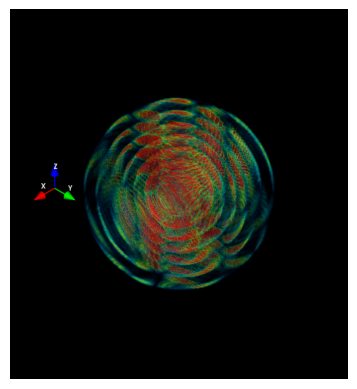}
  \caption{}
  \label{fig:3D.meas}
\end{subfigure}
\begin{subfigure}{.65\textwidth}
  \centering
  \includegraphics[width=1.0\linewidth]{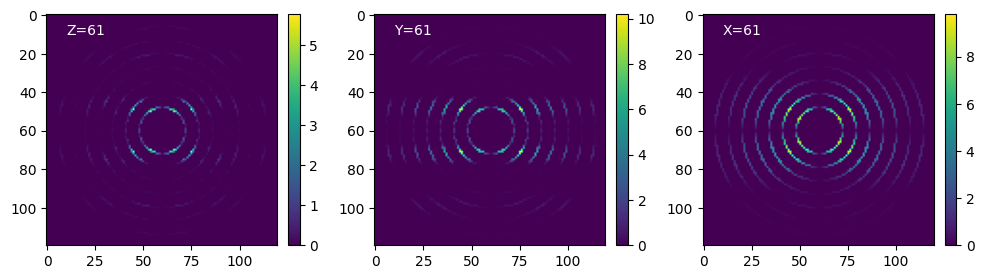}
  \caption{}
  \label{fig:slices.meas}
\end{subfigure}
\caption{(a), (c), (e): Visualization of 3D molecule in physical domain, Fourier domain and measurement on spheres, respectively; (b), (d), (f): exemplary slices of (a), (b), (c) at three axes.}
\label{fig:data}
\end{figure}

To lay the foundation of an image reconstruction method based on the description of the 3D-POT measurement in \eqref{eq:plane_wave_model}, we first define the mathematical image reconstruction problem that follows from this description.
Given that the objective is to reconstruct a 3D image of the molecular orbital, we denote by $D\subset \mathbb{R}^3$ the ``object domain" of the 3D ARPES data. An example of the object (i.e., a molecular orbital) in the domain $D$ is shown in Fig. \ref{fig:data}(a)-(b).
Note that throughout this paper, the presented slices of the orbital or reconstructed orbital in the objective domain are truncated for visualization purposes only.
We denote by $(x_i, y_j, z_l)$ the point in $D$ corresponding to the index $(i,j,l)$, for $(i,j,l)\in \mathbb{I}\equiv \{1,\dots,N_x\}\times \{1,\dots,N_y\}\times \{1,\dots,N_z\}$, where $N_*$ denotes the number of voxels respectively in the $x$, $y$, and $z$ directions. We will denote the total number of voxels $|\mathbb{I}|$ by $N=N_x\times N_y\times N_z$. The number of voxels and their dimensions are indirectly determined by the measurement resolution and maximum measured photoelectron momentum, respectively.
Let $\Dhat$ denote the momentum space, i.e., the Fourier domain, where the ARPES data is measured.  The momentum space is discretized in the same fashion as the object domain, though perhaps with different numbers of pixels/voxels.  We denote by $\vec{k}_{(i,j,l)}$ the point in $\Dhat$ corresponding to the index $(i,j,l)\in \Ibbhat\equiv \{1,\dots,\Nhat_x\}\times \{1,\dots,\Nhat_y\}\times \{1,\dots,\Nhat_z\}$, where $\Nhat_*$ denotes the number of voxels respectively in the $k_x$, $k_y$, and $k_z$ directions in momentum space.  We will denote the total number of voxels $|\Ibbhat|$ by $\Nhat=\Nhat_x\times \Nhat_y\times \Nhat_z$.

We take as the starting point of our algorithm a photon-energy-dependent set of momentum maps. The details on how these momentum maps are acquired differ from one experimental facility to the next; for example, the data structure as retrieved using a toroidal analyser \cite{BROEKMAN20051001} can be subtly different \cite{BRANDSTETTER2021107905} to what would be measured using a time-of-flight momentum microscope \cite{medjanik_direct_2017, keunecke_time-resolved_2020, Wallauer20sci, Neef23nat, baumgartner_ultrafast_2022}. For simplicity and generality, we will therefore assume that we extract from the measurement a two-dimensional intensity distribution $I_{\hbar \omega}$ for the coordinates ($k_i, k_j$) with for $(i,j)\in \Ibbhat\equiv \{1,\dots,\Nhat_x\}\times \{1,\dots,\Nhat_y\}$. Furthermore, we will assume that the measured intensity distribution is already corrected for the polarization factor $\vec{A}\cdot\vec{k}$ (c.f. eq. 1).

Each of these momentum maps is a measurement of the intensity on a hemisphere, which can be completed to a sphere by applying the symmetry expected from a real-valued object, i.e. the molecular orbital.  The momentum maps are therefore in the data domain $\Dhat$ with $k_z >0$ and with radius $r=\|\vec{k}\|$ depending on the photon energy $\hbar \omega$ used for photoemission; these data spheres are denoted  $S_{r}\subset \Dhat$.  The value of $r$ is determined by the experimental parameters and must be mapped to the voxel grid described above. We index the various measurement spheres with the symbol $\rho$ so that the voxels corresponding to the measurement sphere with radius $r_\rho$ are indicated by $\mathbb{S}_{\rho}$. Since a sphere with radius $r$ will run through voxels with an inherent thickness, the measurement sphere will inherit this thickness through the discretization. Complicating matters further is the fact that the measurement device is a planar array, i.e., that $k_z$ is not directly measured. All of these effects of discretization and projection onto a planar grid can be accounted for approximately by an appropriate scaling of the measurements. Accounting for this explicitly in the discussion below adds notational clutter that obscures the main ideas, so we will suppress these specifics in our presentation. Instead, interested readers may refer to the related data repository (ref.~\cite{GROsource}), which includes the script that was used to cast ARPES measurements to $\Dhat$.

A 3D-POT data set containing momentum maps at $m$ photon energies then provides measurement data on $m$ spheres with radius $r_\rho$ for $\rho=1,2,\dots,m$. The measurement spheres taken together are denoted by
\begin{equation}\label{e:meas sphere}
S=\cup_{\rho=1}^m S_{r_\rho}\subset \Dhat\quad\mbox{ and }\quad
\mathbb{S}=\cup_{\rho=1}^m \mathbb{S}_{\rho}\subset\Ibbhat
\end{equation}
is the corresponding collection of voxels.
The equation \eqref{eq:plane_wave_model} can be reformulated in the following form, as presented in, for example, \cite{luke2002optical, luke2019optimization}:
\begin{equation}
|\psihat(\vec{k}_{(i,j,l)})|=
  b_{(i,j,l)}, \ \quad
  (i,j,l)\in \mathbb{S}\,,
\label{eq:basic_phase}
\end{equation}
where
\begin{equation}\label{e:b}
0\leq b_{(i,j,l)}:=\sqrt{\frac{I(\vec{k}_{(i,j,l)})}{|\vec{A}\cdot\vec{k}_{(i,j,l)}|^2
}}\,.
\end{equation}

The problem we are faced with is to find $\psi$ that is consistent with the relationship \eqref{eq:basic_phase}, i.e., find $\psi$ when we know only the amplitude of its Fourier transform on finitely many spheres in momentum space. This is a type of phase problem where the data is sparse and restricted to spheres in the Fourier domain.  The new challenge is to retrieve not only the phase of the measured data but also the phase and Fourier magnitude on the missing voxels $\Ibbhat\setminus\mathbb{S}$. Only by accomplishing this can the complete molecular orbital $\psi$ be calculated.

\subsection{Model and method}
\subsubsection{Model}
The phase retrieval problem has a rich history and many different approaches to its solution. The algorithm we use here is the most prevalent and successful for phase retrieval and is based on a nonconvex {\em feasibility model} \cite{luke2019optimization}.
As explained above, the 3D-POT data records the momentum distribution of a specific molecular orbital on set of spheres. The first step in the feasibility model approach is conceptual: we view the data as specifying a {\em set} of orbitals $\psi$ that {\em could} have produced the data;  i.e. any $\psi$ consistent with the data is acceptable.  We won't belabor the transition in thinking about $\psi$ and $\psihat$ as  functions taking values on $D$ (respectively $\Dhat$) to their discretized versions as vectors in $\mathbb{C}^N$ and $\mathbb{C}^\Nhat$.  This allows us to write $\psihat_{(i,j,l)}$ instead of $\psihat(\vec{k}_{(i,j,l)})$; likewise $\psi_{(i,j,l)}$ corresponds to $\psi(x_i, y_j, z_l)$.
The set of points $\psi$ satisfying \eqref{eq:basic_phase} in this discretized form is thus defined by
\begin{equation}
\text{M} := \{\psi\in {{\mathbb{C}}^{N}}\mid \ |\psihat_{(i,j,l)}|=b_{(i,j,l)}, \  (i,j,l)\in \mathbb{S}\}\,
\label{eq:phase_reformulated}
\end{equation}
where $\mathbb{S}$ is defined by \eqref{e:meas sphere}.
This of course leads to too many possibilities, so we narrow the options down by adding other {\em constraints} or prior knowledge about the orbitals.

\if{
\begin{figure}
    \centering
    \includegraphics[scale=0.5]{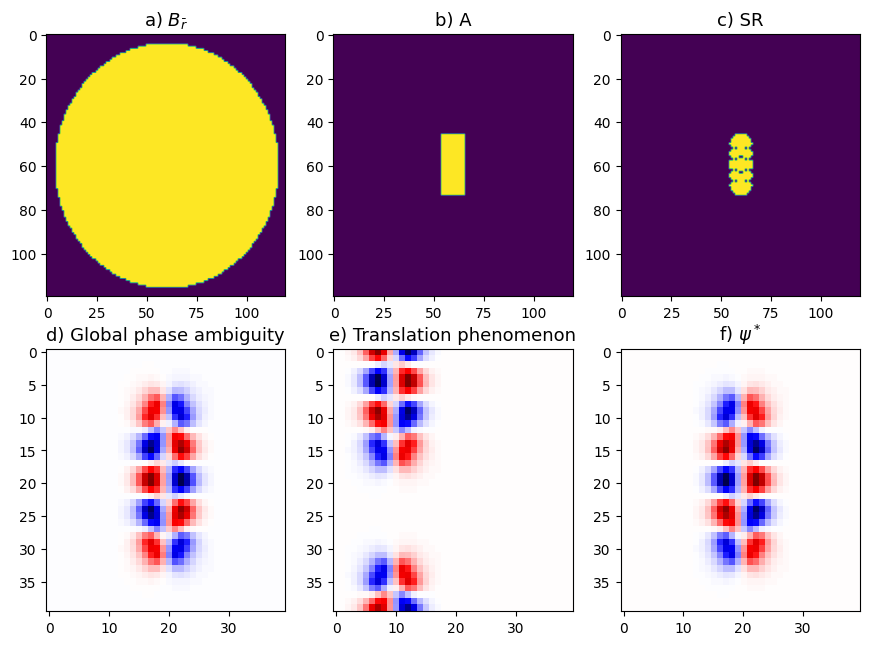}
    \caption{Visualization of (a) the ball $B_{\overline{r}}$ bounding the low pass region in the Fourier domain. The updated Fourier modulus on the dark background will be set to zero, see \eqref{eq:LM}; (b) the support region which limits the updated value only inside the light box and is applied in real space; c)  sparse real constraint removes values as low as zero, allowing us to obtain a tighter support; d) shows a change in sign and e) is an example of a translation, which makes it difficult to use symmetric constraints. f) shows the true molecular orbital.}
\label{fig:constr_visual_and_phase}
\end{figure}
}\fi
\begin{figure}[ht]
   \begin{subfigure}{.3\textwidth}
      \centering
    \includegraphics[width=1\linewidth]{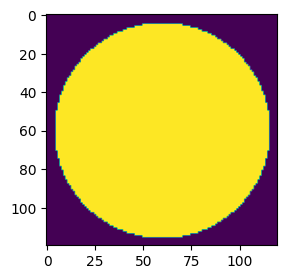}
      \caption{$B_{\bar r}$}
    \end{subfigure}
     \begin{subfigure}{.3\textwidth}
      \centering
    \includegraphics[width=1\linewidth]{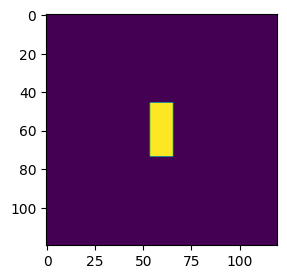}
      \caption{$A$}
    \end{subfigure}
     \begin{subfigure}{.3\textwidth}
      \centering
    \includegraphics[width=1\linewidth]{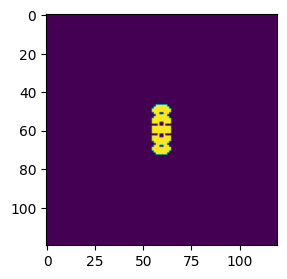}
      \caption{$\SR$}
    \end{subfigure}

     \begin{subfigure}{.3\textwidth}
      \centering
    \includegraphics[width=1\linewidth]{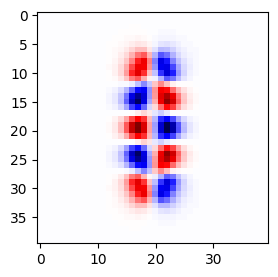}
      \caption{Global phase ambiguity}
    \end{subfigure}
     \begin{subfigure}{.3\textwidth}
      \centering
    \includegraphics[width=1\linewidth]{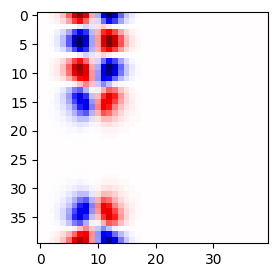}
      \caption{Translation phenomenon}
    \end{subfigure}
     \begin{subfigure}{.3\textwidth}
      \centering
    \includegraphics[width=1\linewidth]{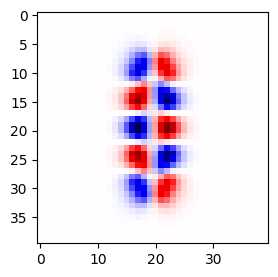}
      \caption{$\psi^*$}
    \end{subfigure}
    \caption{Visualization of (a) the ball $B_{\overline{r}}$ bounding the low pass region in the Fourier domain. The updated Fourier modulus on the dark background will be set to zero, see \eqref{eq:LM}; (b) the support region which limits the updated value only inside the light box and is applied in real space; (c)  sparse real constraint removes values as low as zero, allowing us to obtain a tighter support; (d) shows a change in sign and (e) is an example of a translation, which makes it difficult to use symmetric constraints. (f) shows the true molecular orbital.}
    \label{fig:constr_visual_and_phase}
\end{figure}
The first constraint concerns the experimental resolution in real space. To ensure that the full measurement data can be included in the analysis, the domain $\Dhat$ is chosen to be large enough to fully contain the data for the highest photon energy, which lies on the shell $S_{\overline{r}}$ with the maximum radius $\overline{r}$. Consequently, the corners of $\Dhat$ lie outside the measured area in momentum space. As the momentum distribution in the corners is not constrained by the measurement, and an overestimation of the intensity can lead to high-frequency noise on the reconstructed orbital, we set the intensity outside of the shell $S_{\overline{r}}$ to zero.
This reflects simultaneously the experimental limitations as well as {\em bandwidth} limitations on the types of orbitals which can be successfully recovered. More will be said about this below when we impose {\em symmetry} constraints on the orbital, but imposing such a bandwidth constraint is equivalent to applying a {\em low-pass} filter in momentum space at the determined experimental resolution.
Let $B_{\overline{r}}$ be a ball in momentum space $\Dhat$  that contains all measured spheres $S_j$ ($j=1,\dots,m$) in the Fourier domain. The set of orbitals satisfying the bandwidth constraint is given by
\begin{equation}\label{eq:LM}
    \LF: = \{\psi\in {{\mathbb{C}}^{N}}\mid \ |\psihat_{(i,j,l)}|=0,  \ k_{(i,j,l)} \notin B_{\overline{r}}\}\,.
\end{equation}


The following constraints concern the properties of the molecular orbital in the object domain $D$. Let $s > 0$.
In the object domain $D$, we utilize a sparse-real constraint which is defined as follows:
\begin{equation}
\SR := \{\psi\in {{\mathbb{C}}^{N}}\mid \ \|\psi\|_0 \leq s\mbox{ and }\text{Im}(\psi)=0\},
\label{eq:sparse_set}
\end{equation}
where, Im$(\psi)$ is the imaginary part of the complex vector $\psi$, and $\|\cdot\|_0$ denotes the counting function that counts all non-zero voxels.
We employ the sparse-real constraint since the molecular orbital that we are recovering consists of a small number of positive and negative-valued lobes (e.g., Fig.~\ref{fig:3D.molecule}).
Sparsity constraints were first employed in phase retrieval by Marchesini in  \cite{marchesini2008ab}, and have previously been applied successfully in orbital imaging \cite{jansen2020efficient}.
The sparsity constraint imposed by the restriction on the value of the counting function allows us to impose a movable support constraint on the orbitals;  this is very convenient since we don't know the molecular orbital initially, and the counting function allows the support to evolve during the reconstruction process. This constraint therefore also eliminates the need for a shrink-wrap procedure to fine-tune an a-priori unknown object support \cite{kliuiev_algorithms_2018, Marchesini_x-ray_2003}.

It is not uncommon that the symmetry of a given molecular orbital is known in advance. Based on the properties of the orbital shown in Fig. \ref{fig:data}a and Fig. \ref{fig:data}b, we introduce additional assumptions, namely symmetry and anti-symmetry constraints.
The set of points that are symmetric about the X-axis, anti-symmetric about the Y-axis, and anti-symmetric about the Z-axis is denoted as SYM and is expressed as follows:
\begin{eqnarray}
&&\SYM := \nonumber\\
&&\left\{\psi\in {\mathbb{C}^{N}} \mid \psi_{(i,j,l)}=\psi_{(N_x-i+1,j,l)}=-\psi_{(i,N_y-j+1,l)}=-\psi_{(i,j,N_z-l+1)}\,,\, (i,j,l)\in \mathbb{I}
\right\}.
\label{eq:symmetry_set}
\end{eqnarray}

Imposing a symmetry constraint in the object domain $D$ is made more challenging by the fact that the magnitude of the Fourier transform is invariant under spatial shifts in the object domain. Reconstructions that are consistent with the measured data, and even the sparse-real constraints can appear anywhere in the object domain $D$ (see Fig. \ref{fig:constr_visual_and_phase}e).  While we might see symmetries in the reconstruction with our eyes, to impose these as a constraint we need to align the object with an axis of symmetry (see Fig. \ref{fig:constr_visual_and_phase}d).  We accomplish this by simply applying a conventional, fixed support constraint (as opposed to the sparsity support constraint above) centered at the origin that is symmetric in each of the $x$, $y$, and $z$ directions separately. We emphasize that the size of this support is chosen small enough to ensure proper alignment of the object, but also large enough that it does not affect the shape of the ultimately reconstructed object.  This is shown in Fig. \ref{fig:constr_visual_and_phase}b.
Given a symmetric binary mask $\Omega\in\{0,1\}^{N}$,
the support constraint set is denoted SUPP and given by
\begin{equation}\label{eq:support_constraint}
    \SUPP=\{\psi\in\mathbb{C}^{N}\mid \ \psi\Omega=\psi\}
\end{equation}
We refer to the mask $\Omega$ as the support region. By using the support constraint, we can not only avoid spatial translation of the recovered object, but also speed up the recovery process by reducing the number of indices that need to be estimated.

\begin{figure} 
    \centering
\definecolor{xdxdff}{rgb}{0.49019607843137253,0.49019607843137253,1}
\definecolor{uuuuuu}{rgb}{0.26666666666666666,0.26666666666666666,0.26666666666666666}
\definecolor{ududff}{rgb}{0.30196078431372547,0.30196078431372547,1}
\definecolor{ccqqqq}{rgb}{0.8,0,0}
\begin{tikzpicture}[scale=4,line cap=round,line join=round,>=triangle 45,x=1cm,y=1cm]
\clip(-1.7646394854237981,-0.7929162776221615) rectangle (1.3165930075733998,1.1566635543106134);
\draw[line width=1pt,color=ccqqqq,smooth,samples=100,domain=-1.7646394854237981:1.3165930075733998] plot(\x,{((\x)+1.5)^(2)*(\x)^(2)*((\x)-1)^(2)});
\draw [line width=1pt] (-0.5,-2) circle (1.8775095140394464cm);
\draw [line width=1pt] (-1.5,0)-- (-1.3456012980645164,-0.32369441329444837);
\draw [line width=1pt] (0,0)-- (-0.032061174262565106,-0.18173840422338872);
\draw [line width=1pt] (1,0)-- (0.6406353099138791,-0.5086941745281393);
\draw (-0.9579167963481682,0.98) node[anchor=north west] {$C_1$};
\draw (-0.7954518103537704,0.07) node[anchor=north west] {$C_2$};
\draw (0.9356406266210189,0.15470276999688707) node[anchor=north west] {$a_1$};
\draw (0.48185911401597703,-0.47919078742608284) node[anchor=north west] {$a_2$};
\draw (-0.35,-0.18227201991907974) node[anchor=north west] {$a_2=\widetilde \psi^*$};
\draw (-1.327664695507832,-0.30552131963896784) node[anchor=north west] {$a_2$};
\draw (-1.5629588131548906,0.15) node[anchor=north west] {$a_1$};
\draw (-0.03914928934536733,0.15) node[anchor=north west] {$a_1$};
\draw (-1.41169830895321,-0.05) node[anchor=north west] {gap};
\draw (0.8348002904865652,-0.2) node[anchor=north west] {gap};
\draw (-0.03,-0.05) node[anchor=north west] {gap};
\draw (-0.263238925199709,0.0) node[anchor=north west] {$\psi^*$};
\draw (-0.7058159560120342,-0.25) node[anchor=north west] {$\widetilde E$};
\draw (0.2297582736798424,-0.35) node[anchor=north west] {$\widetilde E$};
\begin{scriptsize}
\draw [fill=ududff] (-0.5,-2) circle (0.5pt);
\draw[color=ududff] (-1.742230521838364,1.2210893246187366) node {$A$};
\draw [line width=0.5pt,dash pattern=on 1pt off 1pt] (-1.3456012980645164,-0.32369441329444837)-- (-0.032061174262565106,-0.18173840422338872);
\draw [line width=0.5pt,dash pattern=on 1pt off 1pt] (-0.032061174262565106,-0.18173840422338872)-- (0.6406353099138791,-0.5086941745281393);
\draw [fill=uuuuuu] (-1.5,0) circle (0.5pt);
\draw [fill=uuuuuu] (-1.3456012980645164,-0.32369441329444837) circle (0.5pt);
\draw [fill=uuuuuu] (0,0) circle (0.5pt);
\draw [fill=uuuuuu] (-0.032061174262565106,-0.18173840422338872) circle (0.5pt);
\draw [fill=uuuuuu] (1,0) circle (0.5pt);
\draw [fill=uuuuuu] (0.6406353099138791,-0.5086941745281393) circle (0.5pt);
\draw [fill=ccqqqq] (-0.08400871459694928,-0.07012345679012469) circle (0.5pt);
\end{scriptsize}
\end{tikzpicture}
    \caption{Example of a non-convex and inconsistent problem involving two sets, $C_1$ and $C_2$. The nearest points between the sets $C_1$ and $C_2$, denoted as $a_1$ and $a_2$ respectively, are shown. Depending on where the starting point $\psi^{(0)}$ is picked, we obtain different solutions $a_1$ and $a_2$.
    }
    \label{fig:consist.inconsist.illustrate}
\end{figure}
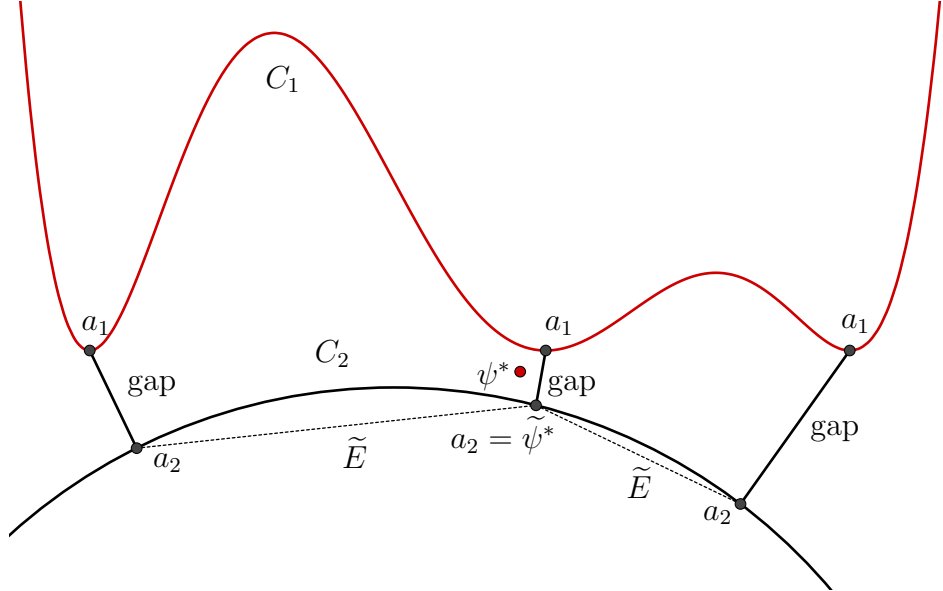

Having defined the qualitative constraint sets above, we can now formulate the feasibility model as finding a point that lies in all these sets:
\begin{equation}
\text{Find}\ \psi\in \text{SYM}\cap \SR \cap \SUPP\cap \LF\cap \M.
\label{eq:intersect_form}
\end{equation}
This is an inconsistent problem because the qualitative constraints are idealized and the intersection above is empty.
From a physical perspective, the most significant impediment is due to the fact that the true molecular orbital in Fourier space is actually non-zero outside the ball $B_{\overline{r}}$. From a numerical perspective, different sparsity parameters $s$ in \eqref{eq:sparse_set} or support regions $\Omega$ in \eqref{eq:support_constraint} may also cause inconsistency, as they introduce hard edges which are suppressed by the low-pass filter LF.  Inconsistent feasibility has been studied in \cite{russell2018quantitative} where the interpretation of ``solutions" to \eqref{eq:intersect_form} is explained in terms of {\em difference vectors} or {\em gaps} of smallest magnitude between successive sets \cite[Lemma 3.2]{russell2018quantitative};  this is depicted in the case of two sets in Fig. \ref{fig:consist.inconsist.illustrate}. Generally, it is found that the size of the gap correlates with the distance to the `true' solution, such that the reconstruction with the smallest gap provides the best approximation.

\subsubsection{Method}
\label{sec:method}
Algorithms based on {\em metric projections} are the most successful methods for finding points where the gap between the sets is smallest \cite{luke2019optimization}.
The Cyclic Projection algorithm (CP) is one of the simplest, and more effective approaches.  Since the focus of this work is not on the algorithms, we will limit ourselves to CP;  if the reconstructions are successful for cyclic projections, they will also be successful for more sophisticated algorithms.
The algorithm CP is given by:
\medskip
    \begin{center}
    		\fcolorbox{black}{white}{\parbox{12.5cm}{\vspace{-0.0in}\textbf{Cyclic Projections:}\\
\noindent Given an initial point $\psi^{(0)}\in{\mathbb{C}^{N}}$,
for $n=1,2,\dots,$ do
\begin{equation}
\psi^{(n)}\in P_{\text{SYM}} P_{\SR} P_{\SUPP}P_{\LF}P_\M \psi^{(n-1)}.
\label{eq:CP}
\end{equation}
		\vspace{-0.0in}}}
	\end{center}
\medskip

Here $\psi^{(n)}$ denotes the reconstructed orbital at the $n$-th iteration and $P_C$ denotes the metric projector onto a set $C$.  The formulas for the individual projectors are given below.
We use character ``$\in$'' instead of ``$=$'' since the projection of a point onto the sets $\M$ and $\SR$ can be multi-valued.

Luke et al. \cite[Theorem 3.2]{russell2018quantitative} proved guarantees on local linear convergence of CP to fixed points, i.e. points $x^*$ with $T_{CP}x^*=x^*$, for non-convex problems and global linear convergence for convex problem under regularity assumptions of the sets.
In short, for our non-convex problem, as long as the starting point is close enough to a fixed point of the algorithm, then the iterates of the algorithm converge linearly to a fixed point.

A projection of $\psi$ onto the set $\M$ with infinite precision arithmetic is computed by \cite[Theorem 4.2]{luke2002optical}
\begin{eqnarray*}
&&\psi^{\M}\in P_{\M}\psi\equiv \\
&&\left\{\mathcal{F}^{-1}\hat v~|~\hat v_{(i,j,l)}\in
\begin{cases}b_{(i,j,l)}\frac{\psihat_{(i,j,l)}}{|\psihat_{(i,j,l)}|}& \mbox{ if } |\psihat_{(i,j,l)}|\neq 0\\
b_{(i,j,l)}e^{\mathbf{i}\theta}\quad  \forall \theta\in[0,2\pi]& \mbox{ else}
\end{cases}
, (i,j,l)\in\mathbb{S}\right\}
\end{eqnarray*}
where $\mathcal{F}^{-1}$ is the discrete inverse Fourier transform and $\mathbf{i}^2=-1$.
As argued in \cite[Corollary 4.3]{luke2002optical},  this formulation is numerically unstable, so instead we approximate the projection by
\begin{equation}
    \psi^{\M} = \mathcal{F}^{-1}(\hat v),
\end{equation}
where $\hat v=(\hat v_{(i,j,l)})_{(i,j,l)\in \mathbb \Ibbhat}$ is defined by
\begin{equation}
\hat v_{(i,j,l)}\in \begin{cases}
\left[1-\frac{|\psihat_{(i,j,l)}|^2+2\epsilon}{\sqrt{\left(|\psihat_{(i,j,l)}|^2+\epsilon\right)^3}}\left(\frac{|\psihat_{(i,j,l)}|^2}{\sqrt{|\psihat_{(i,j,l)}|^2+\epsilon}}-b_{(i,j,l)}\right)\right]\psihat_{(i,j,l)} & \text{if $(i,j,l)\in \mathbb{S}$},\\[20pt]
\psihat_{(i,j,l)} & \text{if $(i,j,l)\notin \mathbb{S}$.}
\end{cases}
\label{eq:PM}
\end{equation}
The first case of \eqref{eq:PM} states that if the triple index $(i,j,l)$ corresponds to voxels that belong to the measured spheres, then the updated magnitude of the Fourier signal must be equal to the measured information.
The parameter $\epsilon$ is introduced to prevent division by zero. The choice of $\epsilon$ depends on
the machine precision.  For our examples $\psihat$ is a complex vector of length of about $10^{6}$ elements,
so with double precision arithmetic, we shouldn't expect better than $10^{-9}$ accuracy for vector-vector multiplications.  To be on the safe side we choose $\epsilon =10^{-8}$ to ensure numerical stability. For comparison, the measurements $b_{(i,j,l)}$ represent photoelectron counts and are on the order of $10^{3}$ or more.
The second case in \eqref{eq:PM} means that if a voxel does not belong to the set of measured spheres, its updated Fourier value at that point is simply the Fourier value of the previous iteration's reconstruction, taken at that index.
This implies that we allow the unmeasured signal to be free as we have no information about it.

The projectors onto the sets LF and SUPP, satisfying the bandwidth constraints in the Fourier domain and the support constraints in the object domain, respectively, are just well known projections onto supports of vectors in the respective domains.  Indeed, recall that the ball $B_{\overline r}$ with radius $\overline r$ represents the measurement bandwidth of the image in momentum space.  The projector $P_{\LF}$ is given by
\begin{equation}\label{eq:P_LM}
     \psi^{\LF} = \mathcal{F}^{-1}(\hat v),
     \quad\mbox{ where }\quad
    \hat v_{(i,j,l)} =\begin{cases}
    \psihat_{(i,j,l)}&\text{ if $(i,j,l)\in B_{\overline r}$}\\
    0 & \text{otherwise.}
    \end{cases}
\end{equation}
The projection $\psi^{\SUPP}$ of a point $\psi\in\mathbb{C}^{N}$ onto the set SUPP (i.e., $\psi^{\SUPP}\in P_\SUPP \psi$) is given by
\begin{equation}\label{eq:P_SUPP}
     \psi^{\SUPP}_{(i,j,l)} =\begin{cases}
    \psi_{(i,j,l)}&\text{if }\Omega_{(i,j,l)}\neq 0,\\
    0 & \text{otherwise.}
    \end{cases}
\end{equation}

The projection of a point $\psi$ onto the sparse-real constraint set,  SR, with some fixed sparsity parameter $s$ is defined by first projecting onto the set of real-valued vectors, and then projecting onto the sparsity constraint by keeping the $s$ components of $\text{Re}(\psi)$ with the greatest absolute values, and then setting the other components to zero. It is important to observe this order of operations since first applying the sparsity constraint, then the real-valued constraint does not yield the {\em nearest} point in the intersection of these two sets to the original point.  This projector is multi-valued whenever there are ties for the $s$th largest component.

A projection $\psi^{\SYM}$ of $\psi$ onto SYM (i.e., $\psi^{\SYM}\in P_\SYM \psi$) is calculated via the following steps:
\begin{enumerate}[label=(\roman*)]
    \item \label{it:symx}Take $u$ as the average of the $\psi$ and its reflection along the X-axis:
    \[
     u =
     \frac{\psi_{([1:N_x],:,:)}+\psi_{([N_x:1],:,:)}}{2}\,.
    \]
    \item \label{it:anti.symy}Take $v$ as the average of $u$ and opposite sign of its reflection along the Y-axis:
    \[
     v =
     \frac{u_{(:,[1:N_y],:)}-u_{(:,[N_y:1],:)}}{2}\,.
    \]
    \item \label{it:anti.symz}Take $\psi^{\SYM}$ as the average of $v$ and the opposite sign of its  reflection along the Z-axis:
    \[
     \psi^{\SYM} =
     \frac{v_{(:,:,[1:N_z])}-v_{(:,:,[N_z:1])}}{2}\,.
    \]
\end{enumerate}
A straight-forward argument shows that this is indeed a metric projector;  we leave this to the reader.

\subsubsection{What to monitor?}\label{subsec:whats.to.monitor}
To decide when to stop the algorithms, we define:
\begin{equation}
\text{change}^{(n)}:= \frac{\|\psi^{(n)}-\psi^{(n-1)}\|}{\|b\|}\,,
\label{eq:change}
\end{equation}
where $\|\psi\|:=\sqrt{\sum_{(i,j,l)\in \mathbb{I}} \psi_{(i,j,l)}^2}$ for all $\psi\in \mathbb{C}^{N}$.
The iterations are terminated once the distance between successive iterates computed by~\eqref{eq:change} reaches a tolerance, for example $\text{TOL}=10^{-14}$. 
The theory \cite[Example 3.6]{russell2018quantitative} establishes that algorithm \eqref{eq:CP} converges locally linearly for this problem;  how long one needs to wait until a region of local linear convergence is found, however, is undetermined.  The simplest solution is therefore  to wait until the expected local convergence behavior is observed.  Figure \ref{fig:convergence.distribution}a) and b) show that the range of possible iteration counts before local linear convergence is observed remains relatively stable under changes in other model parameters, like the number of spheres, or the sparsity parameter.

Iterates of the cyclic projection algorithm \eqref{eq:CP}, and in particular final iterate, always belong to the set SYM but not necessarily to the other sets.
However, we can always look at the intermediate points from the sets SR, SUPP, LF or M simply by storing these intermediate projections. Specifically, if $\overline{\psi}$ is the retrieved orbital obtained after completing the iterative process, the ``shadow" of this point on the set $\M$ is calculated by $ P_{\M}\overline{\psi}$;
the corresponding shadow on the set $\LF$ is computed by $P_{\LF}P_{\M}\overline{\psi}$, and so forth.  Depending on the application, one might prefer one ``shadow" over other possible shadows for the reconstructed orbital. In this paper, we opt for reconstructions on the $\SYM$ set because they do not retain artifacts due to, e.g., measurement noise.
Some practitioners take averages of reconstructions on all sets;  we don't recommend this because the mathematical interpretation could be lost, but in the end, this decision is application dependent.

For the simulated 3D-POT analysis we denote the \emph{truth}, which is the Kohn-Sham molecular orbital that is used for the data simulation, by $\psi^*$.
As mentioned earlier, problem \eqref{eq:intersect_form} is inconsistent, and thus, the point that solves  \eqref{eq:intersect_form} {\em in some sense} is not $\psi^*$, but rather a point close to it;  this ``best numerical solution" is  denoted $\widetilde{\psi}^*$.
There are different ways to define this point.  We explain our approach to this issue in more detail in the next section.  Regardless of how one defines $\widetilde{\psi}^*$, we have the following errors:
\begin{subequations}\label{eq:truths}
\begin{equation}\label{eq:error.with.truth}
E^{(n)}= \frac{1}{2}\min\left\{\left\|\frac{\psi^*}{\|\psi^*\|}\pm \frac{\psi^{(n)}}{\|\psi^{(n)}\|}\right\|
\right\}
\end{equation}
and
\begin{equation}\label{eq:error.with.fixpoint}
\widetilde E^{(n)}= \frac{1}{2}\min\left\{\left\|\frac{\widetilde\psi^*}{\|\widetilde\psi^*\|}\pm\frac{\psi^{(n)}}{\|\psi^{(n)}\|}\right\|
\right\}\,,
\end{equation}
\end{subequations}
where $\psi^{(n)}$ denotes the preferred {\em shadow} of the reconstructed orbital at the $n$-th iteration.  The best numerical solution $\widetilde{\psi}^*$ is constructed so that it  belongs to the same set.
The error is the minimum of either the sum or the difference of the reconstruction and the reference ``truth" to account for the unavoidable global phase ambiguity, e.g., the change in sign, shown in Fig. \ref{fig:constr_visual_and_phase}(d).
The formula \eqref{eq:error.with.truth} is the error between the reconstruction and actual orbital. This metric helps us to estimate reliability of the model. 
On the other hand, formula \eqref{eq:error.with.fixpoint} tells us how far the algorithm is from the best numerical truth to the model problem.  Thus, the model error \eqref{eq:error.with.fixpoint} could be small, while the physical error \eqref{eq:error.with.truth} is large;  in this case,  it may be necessary to consider adding additional constraints or modifying the model.
It is easy to see that $0\leq E^{(n)},\widetilde E^{(n)}\leq 1$ for all $n\in\mathbb{N}$. Moreover, when the model fits perfectly with the truth, i.e, $\psi^*=\widetilde \psi^*$, we get $E^{(n)}=\widetilde E^{(n)}$.

The accuracy of the reconstructed orbital is also influenced by the initial starting point, given the non-convex nature of the problem. While in simulation it is trivial to compare the reconstructed orbital and the truth, this is not possible in experiment. Thus, there is a challenge in selecting the best reconstruction among the multiple possibilities.
To address this challenge, we monitor the sum of the gaps between the sets at any given iterate:
   \begin{eqnarray}\label{e:gap}
    &&\text{gap}^{(n)}:=\\
    &&\frac{\text{gap}^{(n)}_{\SYM-\M}+ \text{gap}^{(n)}_{M-\LF}+\text{gap}^{(n)}_{\LF-\SUPP}+\text{gap}^{(n)}_{\SUPP-\SR}+\text{gap}^{(n)}_{\SR-\SYM}}{\|b\|},
    \nonumber
\end{eqnarray}
where
\begin{equation*}
    \begin{array}{rcl}
             & \text{gap}^{(n)}_{\SYM-\M} &:= \|\psi^{(n)}-P_{\M}\psi^{(n)}\|,\\
         & \text{gap}^{(n)}_{M-\LF}&:=\|P_{\M}\psi^{(n)}-P_{\LF}P_{\M}\psi^{(n)}\|\,, \\
         & \text{gap}^{(n)}_{\LF-\SUPP} &:= \|P_{\LF}P_{\M}\psi^{(n)}-P_{\SUPP}P_{\LF}P_{\M}\psi^{(n)}\|\,, \\
         & \text{gap}^{(n)}_{\SUPP-\SR} &:= \|P_{\SUPP}P_{\LF}P_{\M}\psi^{(n)} - P_{\SR}P_{\SUPP}P_{\LF}P_{\M}\psi^{(n)}\|\,,\\
         & \text{gap}^{(n)}_{\SR-\SYM} &:=\\
         &&\|P_{\SR}P_{\SUPP}P_{\LF}P_{\M}\psi^{(n)}-P_{\SYM}P_{\SR}P_{\SUPP}P_{\LF}P_{\M}\psi^{(n)}\|.
    \end{array}
\end{equation*}
When the algorithm converges, the gap remains constant but nonzero. For a sufficient number of trials, the reconstruction with the smallest gap corresponds to the point where the different constraints are nearest to each other. If the model of the molecular orbital is reliable, then it is likely that this point provides the best approximation of the true solution (see Fig . \ref{fig:consist.inconsist.illustrate}). However, a smaller gap does not necessarily have to imply a smaller error. In our numerical experiments, we will investigate to what extent the gap can be used as a measure for the reconstruction accuracy.

\section{Numerical results}
\label{sec: Numerical results}
The implementation of algorithms described in this section can be downloaded from the link \cite{GROsource}.

Starting with the highest occupied molecular orbital of an isolated pentacene molecule as calculated by density functional theory (Fig.~\ref{fig:3D.molecule}), we generated 3D-POT data for up to 56 photoelectron kinetic energies in the range up to 110 eV such that the spheres are equidistantly spaced in momentum. The individual momentum maps span a momentum range of $11.7 \times 11.7$~\AA{}\textsuperscript{-2} with $N_x=N_y=120$ pixels (although our simulated measurement data only covers the central part of the image up to a $=5.5$~\AA{}\textsuperscript{-1} radius). From these momentum maps, we calculate the measured intensities $I(\vec{k}_{(i',j',l')})$ in the data domain $\Dhat$, which has dimensions $N_x=N_y=N_z=120$. Each voxel in $\Dhat$ therefore has a volume $0.098 \times 0.098 \times 0.098$~\AA{}\textsuperscript{-3}. Since the orbital is real-valued, we set the intensities $I(-\vec{k})=I(\vec{k})$.
To define the symmetric support region $\Omega$ in the object domain $D$, we chose a shape of $28 \times 12 \times 10$ voxels ($= 15 \times 6.4 \times 5.4$~\AA{}\textsuperscript{3}) located at the center of the image. The radius of the low-pass filter $B_{\overline{r}}$ is set to $\overline{r}=56$ ($=5.5$~\AA{}\textsuperscript{-1}), as shown in Fig. \ref{fig:constr_visual_and_phase}a for the 2D case;  as claimed, the bandwidth contains all the spheres for which we simulated measurements to perform the reconstructions.
We start with ideal data, generated using a fixed total number of $10^{12}$ detected photoelectrons with constant, well-calibrated vector potential $A$ (i.e., the probe light intensity).

\subsection{Convergence}

\if{
\begin{figure}[bht]
\centering\includegraphics[width=\textwidth]{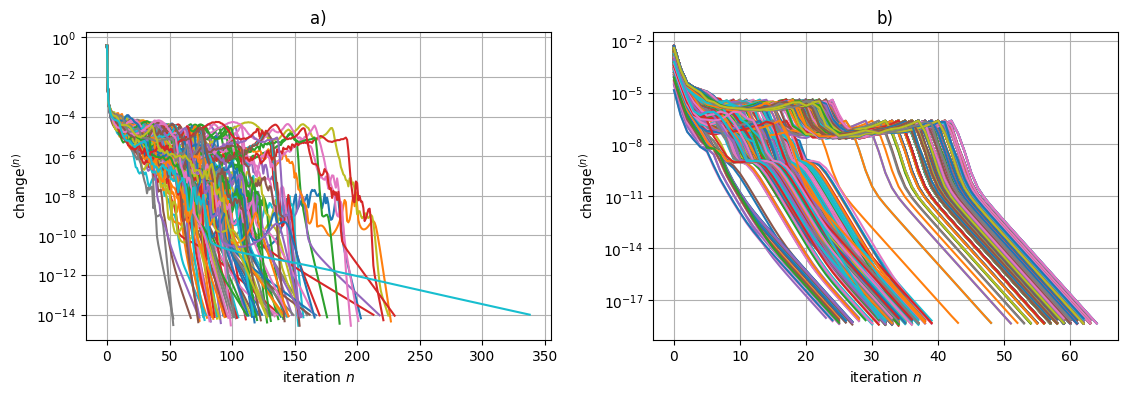}
    \caption{Algorithm performance of locally and globally random initializations.  (a) Iterate differences of 100 globally random initializations, $m=56$ and $s=3360$.  
    (b) Iterate differences of 1000 locally random initializations.  
    }\label{fig:convergence.distribution}
\end{figure}
}\fi
\begin{figure}[ht]
    \begin{subfigure}{.45\textwidth}
      \centering
      \includegraphics[width=1\linewidth]{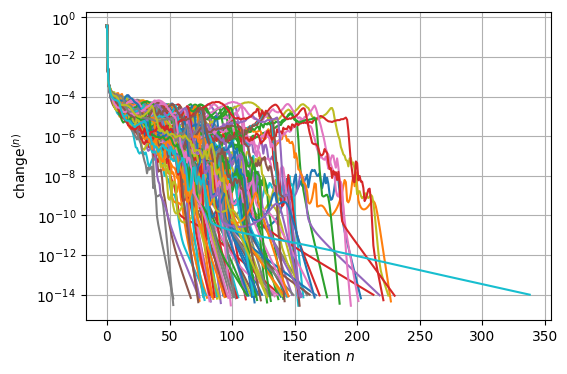}
      \caption{}
    \end{subfigure}
    \begin{subfigure}{.45\textwidth}
      \centering
      \includegraphics[width=1\linewidth]{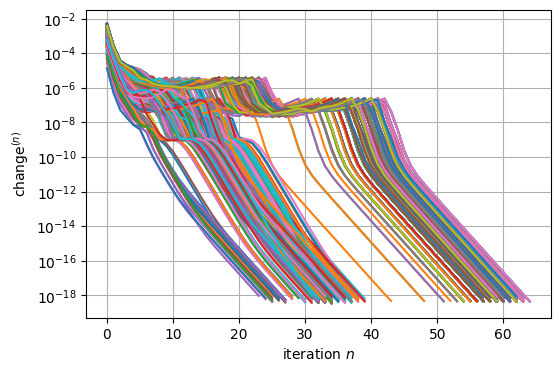}
      \caption{}
    \end{subfigure}
    \caption{Algorithm performance of locally and globally random initialization. (a) Iterate differences of 100 globally random initializations   (b) Iterate differences of 1000 locally random initializations.
    }
    \label{fig:convergence.distribution}
\end{figure}
In Fig.\ref{fig:convergence.distribution}, we investigate the convergence behavior under various scenarios. 
Frame a) shows 100 globally random instances (i.e., the starting points are chosen randomly over the entire domain space) of the algorithm behavior with a fixed sparsity value $s=3360$ and fixed data including all $m=56$ measured spheres.
All instances exhibit local linear convergence, as expected.  The variation of the slopes of the graphs in this frame indicates considerable heterogeneity of local minima.  In our experiments, we observed no correlation between the rate of convergence of the algorithm and the quality of the local minima.
It is interesting to observe in frame b) the convergence plots of 1000 locally random instances using data $m=56$ and $s=3360$, i.e., instances whose starting points are randomly chosen in a small neighborhood of a fixed point. Here, we specifically select the neighborhood around the starting point of the best numerical reconstruction, $\widetilde{\psi^*}$.
Most of the graphs have the same slope in the final iterations, indicating a constant of linear convergence concentrated around the value $\approx 0.38$, and its variance is only $6.12\times10^{-5}$.
This indicates that the geometry of the problem around this local minimum is not wildly varying;
a mathematical statement about why this should be the case is beyond the scope of this work.
As argued in \cite{ACL16}, under the assumption of Q-linear convergence, the rate of convergence $r$  of each instance can be estimated empirically from the observed local numerical behavior.
Here, we observe that the mean of $r$ is approximately $0.28$ with a small variance of about $1.85\times 10^{-5}$.
Moreover, the average error $\widetilde E$ over 1000 reconstructions is $3.28\times 10^{-12}$ with a variance of about $4.66\times 10^{-33}$, suggesting that the fixed point with the smallest gap is an isolated point, which further bolsters the hypothesis that numerical convergence is Q-linear.

In order to benchmark the reconstruction depending on various parameters such as the sparsity parameter, the number of photon energies and the intensity calibration, we now define the best numerical solution $\widetilde \psi^*$, which may be referred to as the ideal reconstructed orbital for the inconsistent problem \eqref{eq:intersect_form}.
This is constructed with  the fixed point of algorithm \eqref{eq:CP} obtained using data from 56 spheres of ideal data with the setting $s=3360$ in Fig. \ref{fig:convergence.distribution}a);  note that we constructed $\widetilde \psi^*$ so that it belongs to the set SYM, and from $100$ globally random initializations it was chosen as the solution that achieved the smallest gap. The errors that we report are the distances between points computed by the algorithm in the set SYM. The distance between the truth and the best numerical solution, i.e., $\frac{1}{2}\min\{\big\|\frac{\psi^*}{\|\psi^*\|}\pm\frac{\widetilde \psi^*}{\|\widetilde \psi^*\|}\big\|\}$, is $7.398\%$.

In the following subsection (\ref{subsec:initial.results}), we present exemplary numerical results that demonstrate the successful reconstruction of the orbital using as little as $m=4$ photon energies. In subsection \ref{subsec:factors.impact}, we then conduct a more comprehensive investigation of the effects of the sparsity parameter $s$, the number of spheres $m$, the starting points, and calibration errors (defined below), exploring their impact on the reconstructions.

\subsection{Example: 4 photon energies}\label{subsec:initial.results}
\if{
\begin{figure}[htb]
    \centering
    \includegraphics[width=\textwidth]{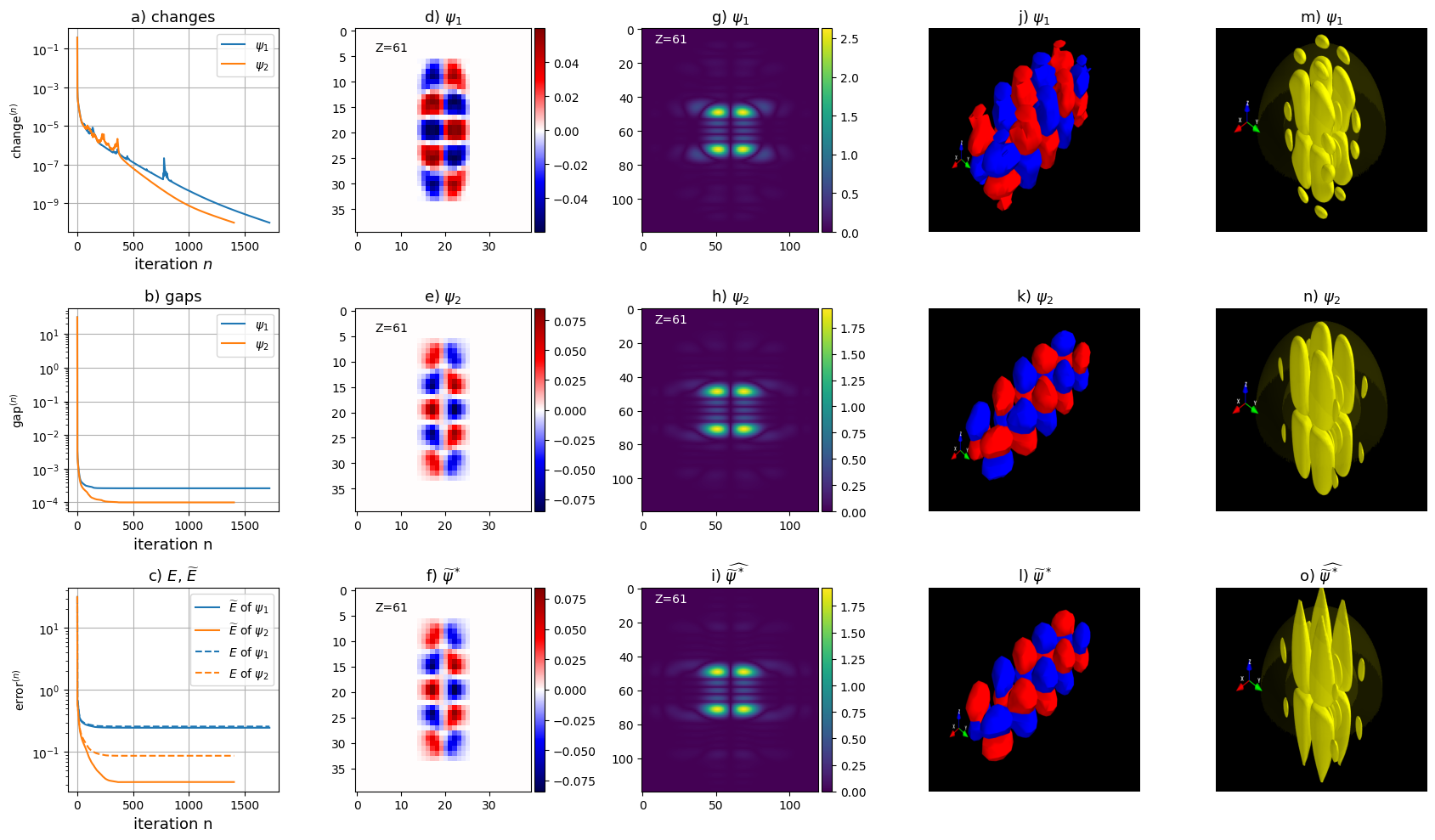}
    \caption{Comparison of retrieval results $\psi_1$, $\psi_2$ at two different starting points with the best numerical solution $\widetilde \psi^*$ using ideal data on  $m = 4$ spheres. The sparsity is set to $s = 3360$. The reconstructions have been normalized by dividing them by their norms for the purpose of visualization.
}
    \label{fig:gaps.and.error}
\end{figure}
}\fi
\begin{figure}
\begin{subfigure}{.19\textwidth}
      \centering
      \includegraphics[width=1.\linewidth]{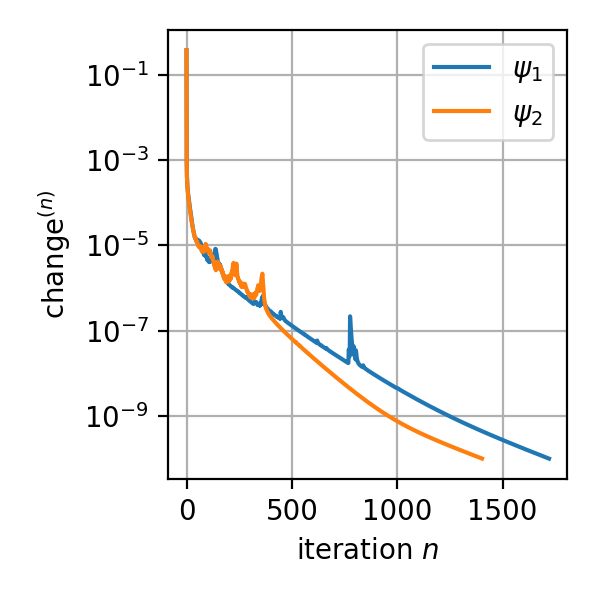}
      \caption{changes}
    \end{subfigure}
    \begin{subfigure}{.19\textwidth}
      \centering
      \includegraphics[width=1.\linewidth]{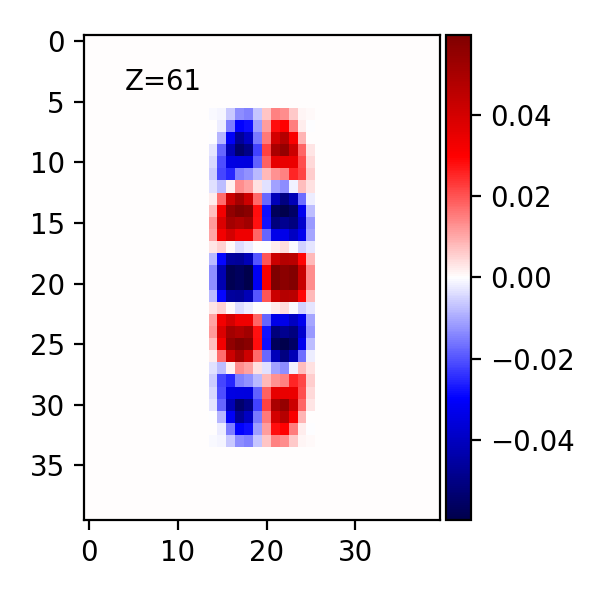}
      \caption{$\psi_1$}
    \end{subfigure}
    \begin{subfigure}{0.19\textwidth}
      \centering
      \includegraphics[width=1.\linewidth]{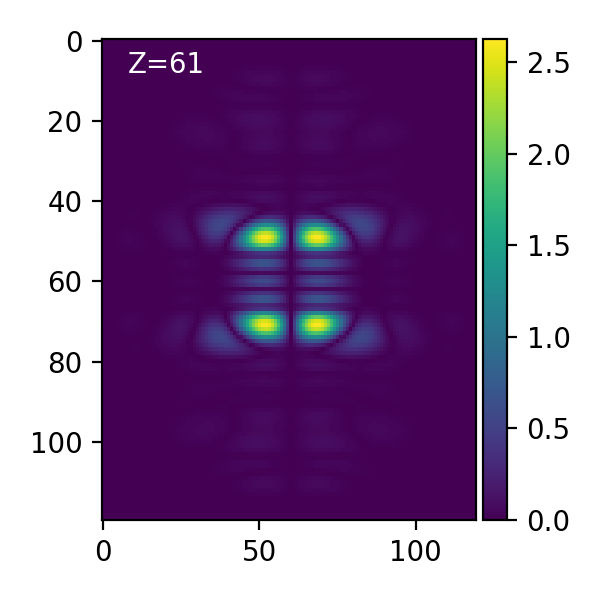}
      \caption{$\widehat{\psi_1}$}
    \end{subfigure}
    \begin{subfigure}{.19\textwidth}
      \centering
      \includegraphics[width=1.\linewidth]{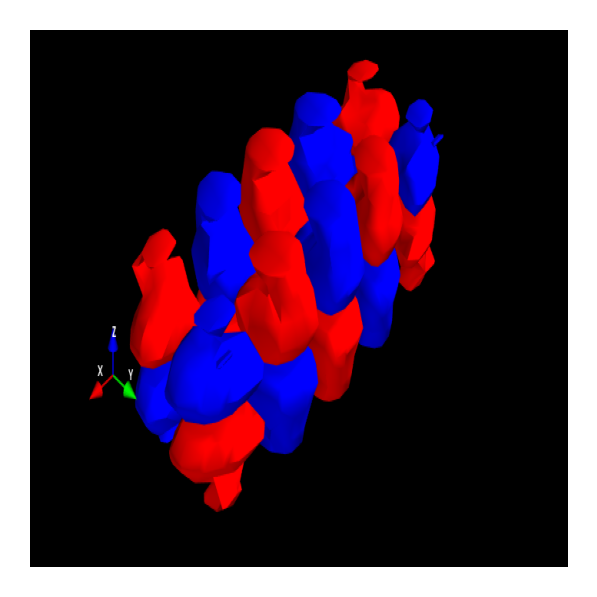}
      \caption{$\psi_1$}
    \end{subfigure}
    \begin{subfigure}{.19\textwidth}
      \centering
      \includegraphics[width=1.\linewidth]{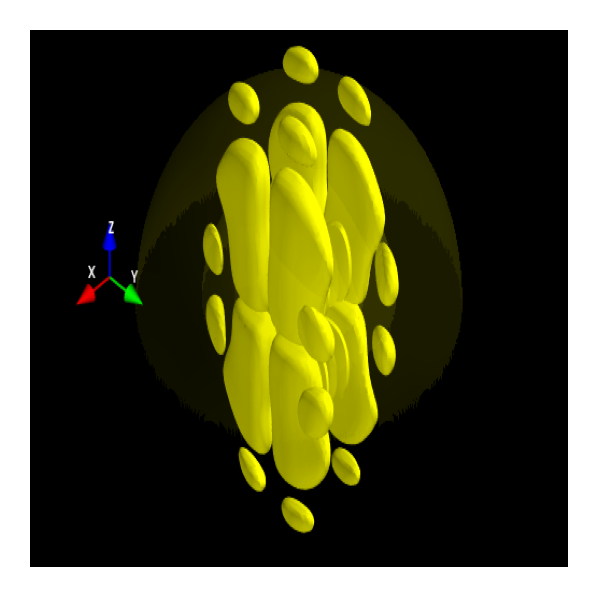}
      \caption{$\widehat\psi_1$}
    \end{subfigure}

    \begin{subfigure}{.19\textwidth}
      \centering
      \includegraphics[width=1.\linewidth]{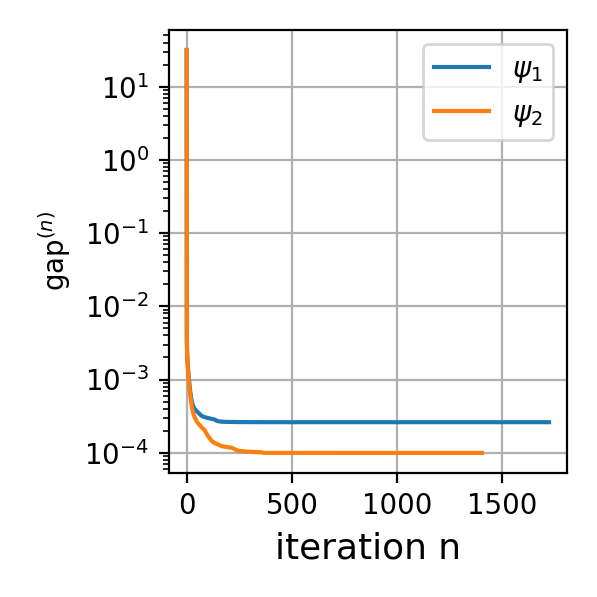}
      \caption{gaps}
    \end{subfigure}
    \begin{subfigure}{.19\textwidth}
      \centering
      \includegraphics[width=1.\linewidth]{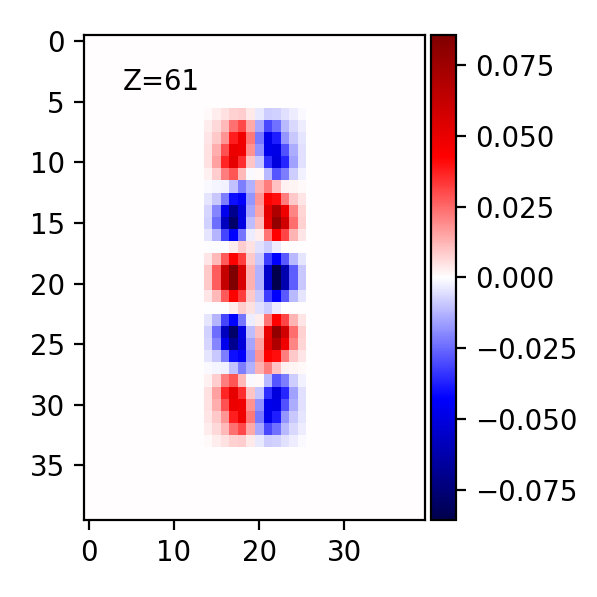}
      \caption{$\psi_2$}
    \end{subfigure}
    \begin{subfigure}{.19\textwidth}
      \centering
      \includegraphics[width=1.\linewidth]{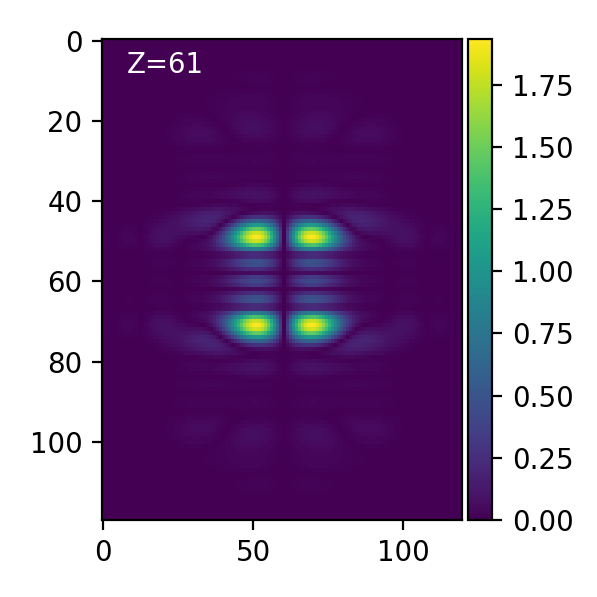}
      \caption{$\widehat\psi_2$}
    \end{subfigure}
    \begin{subfigure}{.19\textwidth}
      \centering
      \includegraphics[width=1.\linewidth]{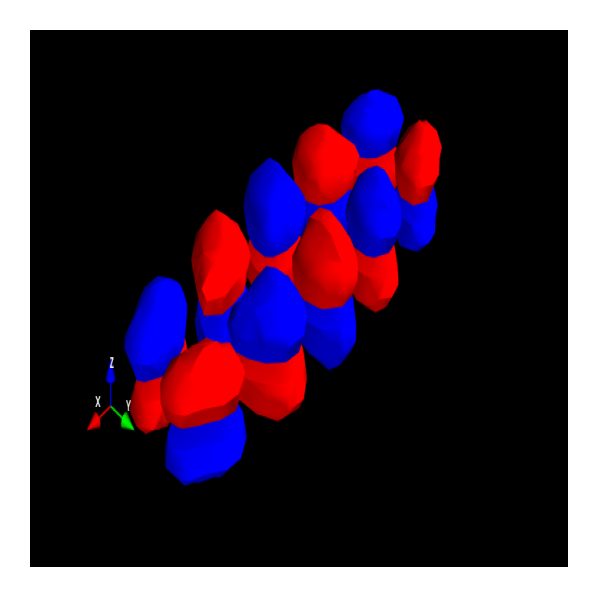}
      \caption{$\psi_2$}
    \end{subfigure}
    \begin{subfigure}{.19\textwidth}
      \centering
      \includegraphics[width=1.\linewidth]{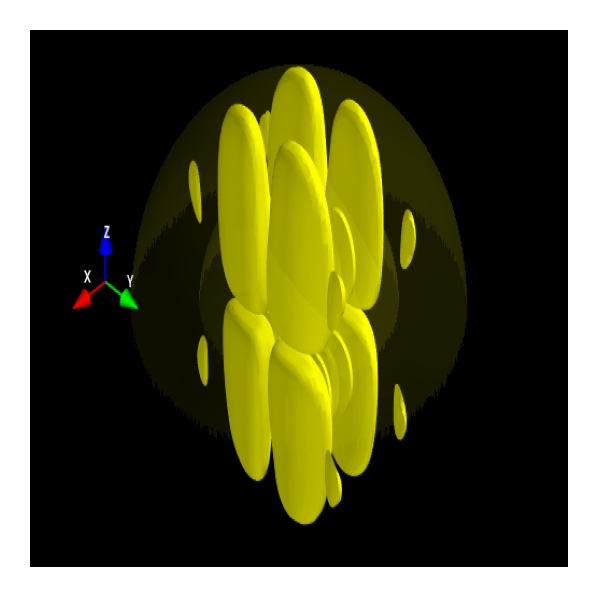}
      \caption{$\widehat\psi_2$}
    \end{subfigure}

    \begin{subfigure}{.19\textwidth}
      \centering
      \includegraphics[width=1.\linewidth]{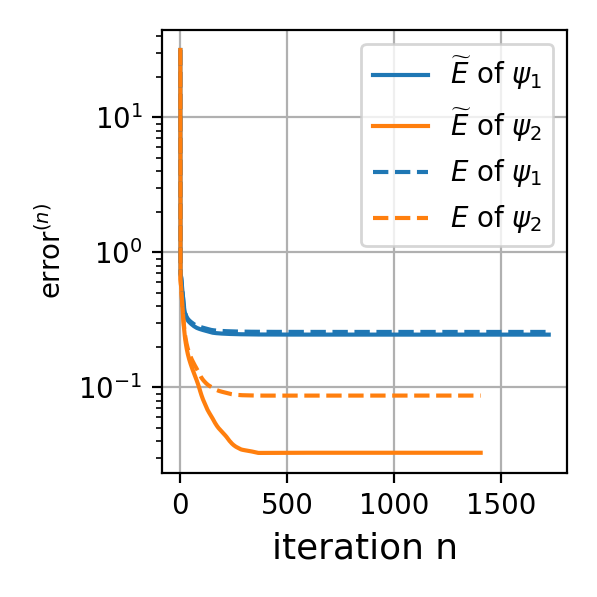}
      \caption{errors}
    \end{subfigure}
    \begin{subfigure}{.19\textwidth}
      \centering
      \includegraphics[width=1.\linewidth]{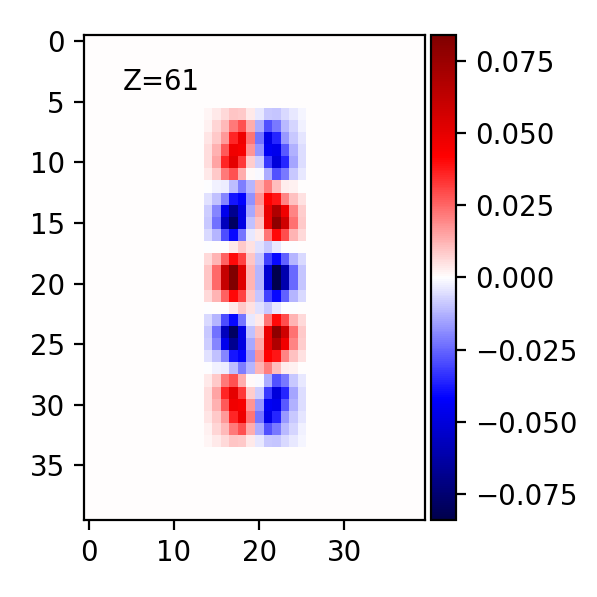}
      \caption{$\widetilde\psi^*$}
    \end{subfigure}
    \begin{subfigure}{.19\textwidth}
      \centering
      \includegraphics[width=1.\linewidth]{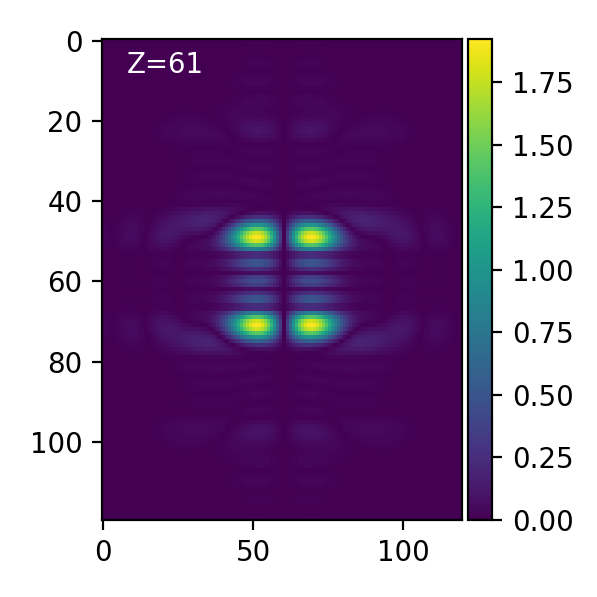}
      \caption{$\widehat{\widetilde\psi^*}$}
    \end{subfigure}
    \begin{subfigure}{.19\textwidth}
      \centering
      \includegraphics[width=1.\linewidth]{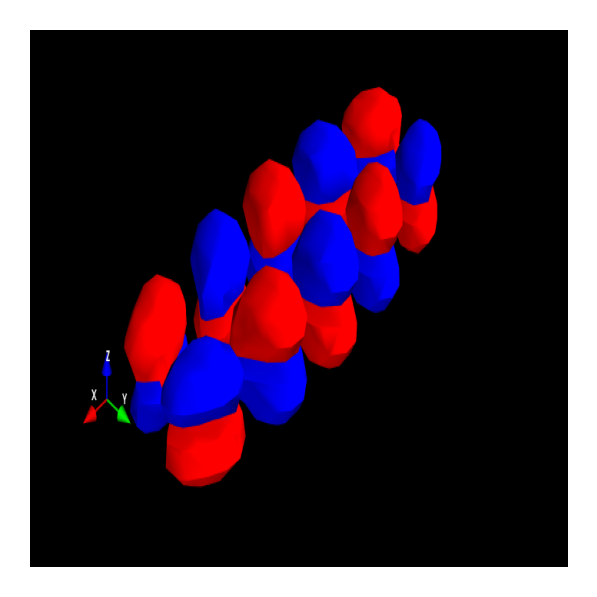}
      \caption{$\widetilde\psi^*$}
    \end{subfigure}
    \begin{subfigure}{.19\textwidth}
      \centering
      \includegraphics[width=1.\linewidth]{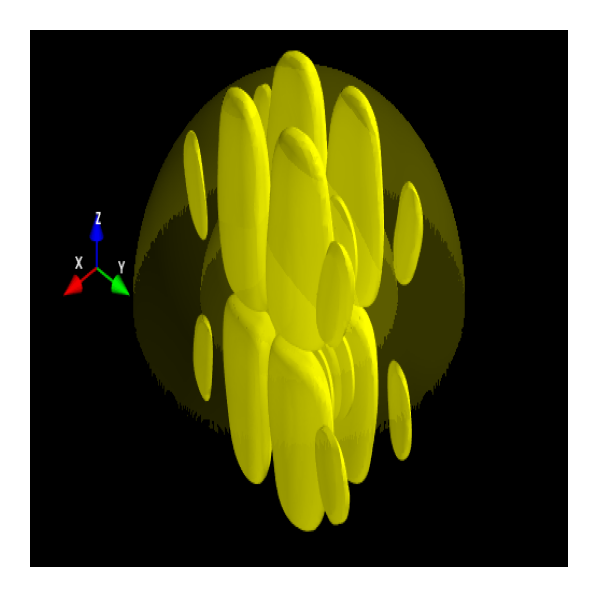}
      \caption{$\widehat{\widetilde\psi^*}$}
    \end{subfigure}
\caption{Comparison of retrieval results $\psi_1$, $\psi_2$ at two different starting points with the best numerical solution $\widetilde \psi^*$ using ideal data on  $m = 4$ spheres. The sparsity is set to $s = 3360$. The reconstructions have been normalized by dividing them by their norms for the purpose of visualization. }
\label{fig:gaps.and.error}
\end{figure}
Figure \ref{fig:gaps.and.error} shows the results of typical reconstructions using $m=4$ spheres (radii: $1.9$, $3.2$, $3.9$ and $4.6$~\AA\textsuperscript{-1}, photoelectron kinetic energies $13.1$, $39.6$, $58.2$, and $80.3$~eV, respectively) of ideal data and sparsity $s=3360$ at two different starting points, in comparison to the ideal reconstructed orbital.
Here, one point, denoted $\psi_1$, achieves an error of $\widetilde{E}=0.25$, while another point, denoted $\psi_2$, achieves $\widetilde{E}=0.04$.
These reconstructed orbitals are selected from a pool of $100$ experiments with varying starting points.
The reconstruction $\psi_2$ is the result with the smallest gap among the $100$ random initializations;  $\psi_1$ is just a randomly selected trial.

Frame (a) of Fig. \ref{fig:gaps.and.error} shows that the algorithm is converging at a locally linear rate on termination for both instances, as expected.
In frames (f) and (k), it is clear that the quality of the reconstructed orbital is affected by the choice of starting points, as measured by the gap and the error, respectively.
The fact that the gaps and errors appear unchanged from iteration 500 onward does not necessarily indicate that the solutions have stopped changing, as indicated in frame a).
Our objective is to find a point whose corresponding {\em difference vectors} are as small as possible;  by this reasoning, the better solution should be $\psi_2$ (orange) as it corresponds to the reconstruction with the smallest gap.
Indeed, visual inspection of the constant height (z) maps in frames (b),(g), (l),the constant $k_z$ maps in frames (c), (h), (m) and the isosurfaces in frames (d), (i), (n), (e), (j), (o) confirms that $\psi_2$ is a better approximation of the molecular orbital.
Further experimental results regarding the relationship between the gap and error can be found in the next Subsection. 

\subsection{Reconstruction accuracy in experiment}\label{subsec:factors.impact}
In subsection \ref{subsec:initial.results} we have seen that, although good orbital reconstructions can be achieved using the CP algorithm for 3D-POT, also but not all random starting points yield the same reconstruction accuracy. In order to investigate how the most optimal results can be achieved, it is first useful to characterize the impact of the sparsity parameter $s$ on the reconstruction result since the sparsity parameter is one of the few control parameters that the user can easily adjust. In this context, also the size (whether larger or smaller) of the support or low-pass filter LF may impact the quality of the reconstruction. In particular, without a low-pass constraint at all, the algorithm tends towards orbitals with un-physically strong high-frequency components. However, a detailed discussion of the effects of the low-pass and support constraints is beyond the scope of the present paper. 

\if{
\begin{figure}
\centering
    \includegraphics[width=\textwidth]{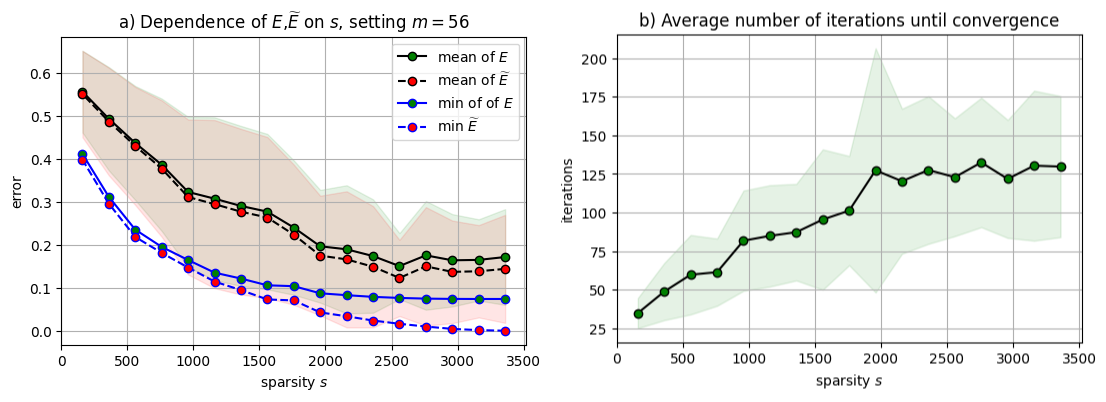}
    \caption{Dependence of a) errors $E$ and $\widetilde{E}$ on the sparsity parameter $s$, and b) the average number of iterations needed to achieve convergence using ideal data. The data points indicate the mean error and minimum achieved error (out of 100 trials), while the shaded regions indicate the standard deviation: in frame a) the green shaded region is the standard deviation of $E$, the pink-shaded region is the standard deviation of $\widetilde{E}$ and the brown shading indicates the overlap. How these results were obtained is described in subsection \ref{subsec:factors.impact}.}
    \label{fig:s.m.and.error}
\end{figure}
}\fi
\begin{figure}
    \begin{subfigure}{.45\textwidth}
      \centering
      \includegraphics[width=1.\linewidth]{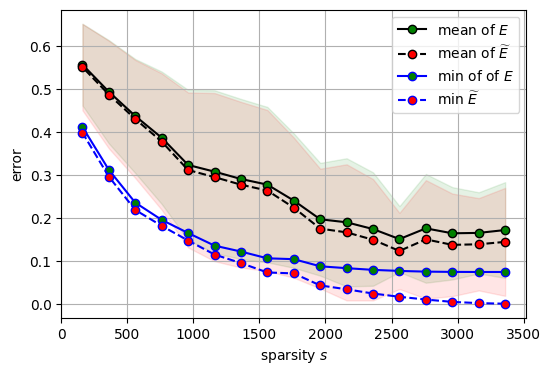}
      \caption{}
    \end{subfigure}
    \begin{subfigure}{.45\textwidth}
      \centering
      \includegraphics[width=1.\linewidth]{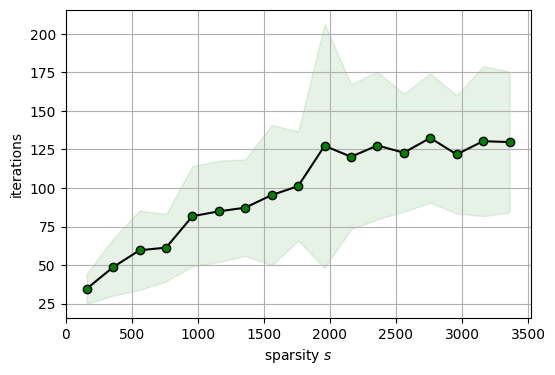}
      \caption{}
    \end{subfigure}
    \caption{Dependence of (a) errors $E$ and $\widetilde{E}$ on the sparsity parameter $s$, and (b) the average number of iterations needed to achieve convergence using ideal data, setting $m=56$. The data points indicate the mean error and minimum achieved error (out of 100 trials), while the shaded regions indicate the standard deviation: in frame a) the green shaded region is the standard deviation of $E$, the pink-shaded region is the standard deviation of $\widetilde{E}$ and the brown shading indicates the overlap. How these results were obtained is described in subsection \ref{subsec:factors.impact}.}
    \label{fig:s.m.and.error}
\end{figure}
    \subsubsection{The sparsity parameter $s$}\label{item:sparse.investigate}
    In Fig. \ref{fig:s.m.and.error}a), we present the dependence of errors $E$ and $\widetilde E$ on the sparsity parameter $s$ for a dataset consisting of $m=56$ input spheres of ideal data.
    Each $s$ value underwent testing through a total of $100$ globally random trials [i.e., we use the same reconstructions as in Fig. \ref{fig:convergence.distribution}a)].
    We then calculated the averages of $E$ and $\widetilde{E}$ across these starting points.
    We also identify the minimum values of $E$ and $\widetilde E$ among the $100$ trials for each value of $s$.
    Here, the minimum achieved error provides an indication of the reconstruction quality that can be achieved, while the mean and spread show that a significant number of trials is necessary to achieve the best result.

    Figure \ref{fig:s.m.and.error}a) shows a trend that \if{is opposite to the result from Fig. \ref{fig:convergence.distribution}b: where }\fi the number of required iterations increases as the number of nonzero voxels $s$ increases, the error decreases as $s$ increases. This holds for both the mean and the minimum errors of $E$ and $\widetilde{E}$. The data shows that a wide spread of reconstruction errors is to be expected for all $s$. Therefore, it is necessary to perform reconstructions with multiple globally random instantiations and select the best one. Having established that larger values for $s$ generally lead to better reconstruction results, with diminishing returns for $s>2500$, we choose to set the sparsity parameter to $s=3360$ for the remainder of our investigation.

    Figure \ref{fig:s.m.and.error}b) displays the average number of iterations until convergence over 100 trials, corresponding to the data presented in Fig. \ref{fig:s.m.and.error}a) for each sparsity level $s$. A smaller value of $s$ results in a faster reconstruction process and a smaller spread.
    The average number of iterations remains the same when $s$ is large, but there is a larger relative variation.

    \subsubsection{Number of spheres}\label{item:m.investigate}
    As the number of photon energies and thereby measured spheres increases, the algorithm is provided with more information and needs to fill fewer voxels. In Fig. \ref{fig:m.error.and.gap}a), we run a total of $500$ reconstructions using ideal data with a fixed sparsity parameter $s=3360$ for five various settings of $m$: $4, 7, 13, 26, 56$. The $m=4$ data includes the same spheres as described in subsection~\ref{subsec:initial.results}, the $m=7$ data covers radii of $1.2$ to $5.2$~\AA\textsuperscript{-1} ($5.2$,  $13.1$,  $24.6$,  $39.6$, $58.2$,  $80.3$, and $106.1$~eV kinetic energy), and the $m = 13,26,56$ data sets cover radii of $0.5$ to $5.4$~\AA\textsuperscript{-1}.
    Again, each value of $m$ was tested with $100$ globally random instances.

    Similarly to Fig. \ref{fig:s.m.and.error}a), we measure the mean values of $E$ and $\widetilde E$, along with their spreads and respective minimum values, and find that the mean errors decrease as the number of measurements increases.
    Moreover, the decrease in spreads indicates that the errors exhibit less variation when more data is used.
    Also as expected, the error measured against the best numerical solution,  $\widetilde E$, goes to zero as the number of shells increases to $56$.

    Rather surprisingly, whereas the mean $E$ and $\widetilde E$ increases significantly as the number of spheres is reduced, the minimum error increases much less. For all numbers of spheres, the minima of $E$ and $\widetilde E$ stay below $0.1$.
    This observation leads to the conclusion that a dense input is not necessary; using $13, 7$ or even $4$ shells can still yield a close approximation of the orbital as long as a sufficient number of trials is performed. Finally, we note that the simulated data for $m \geq 13$ also includes photoelectron kinetic energies below 5~eV which are often inaccessible in experiment, e.g., due to strong final-state effects or the presence of a strong inner potential \cite{Damascelli_2004}. To verify that the inclusion of these shells in the reconstruction does not affect our conclusions, we have also performed reconstructions where we exclude shells for which the photoelectron kinetic energy is less than 14~eV. From this analysis, we find that the exclusion of these shells leads to an increase of the best feasible error of less than $10^{-3}$ from $E=0.07398$ to $E=0.07415$.
    We thus conclude that the the lower kinetic energy shells do not affect the reconstruction strongly.

    In all of these numerical experiments, we chose the photon energies such that the radii of the corresponding spheres are evenly spaced in momentum (for example, see in Fig. \ref{fig:data}e) and f) for the case of 7 spheres). It is possible that the energy spacing can be adapted to the specific molecular orbital of interest to provide better reconstruction results, however this falls outside of the scope of the present study.

    In subsection \ref{subsec:initial.results}, we have observed an example where examining the gap can offer insights for selecting a reconstruction when both the ground truth and the ideal reconstruction are unknown. Using the full dataset of $500$ numerical reconstructions with different $m$, we can investigate more quantitatively to what extent the gap can be used as an indication for the error $E$ (or $\widetilde{E}$).
    In Fig. \ref{fig:m.error.and.gap}b), we depict the dependence of the gap and $\widetilde{E}$ for each random instance and input data $m$.
    First, we note a clear trend in the data, where a larger error $\widetilde{E}$ generally corresponds to a larger gap. Surprisingly, this trend is highly similar between datasets with different $m$. We emphasize that this match between different $m$ is not expected to be general, but in this case it further supports that there is a direct correlation between the error $\widetilde{E}$ and the gap.
    However, we also observe that the correlation between the gap and $\widetilde{E}$ weakens as the amount of data decreases;  this is particularly true for $m=4$.
    This is because the problem becomes more ill-defined and, therefore, more non-convex.
    However, even for $m=4$ the smallest gaps do result in the smallest $\widetilde{E}$ as claimed.
\if{
\begin{figure}
    \centering
    \includegraphics[width=\textwidth]{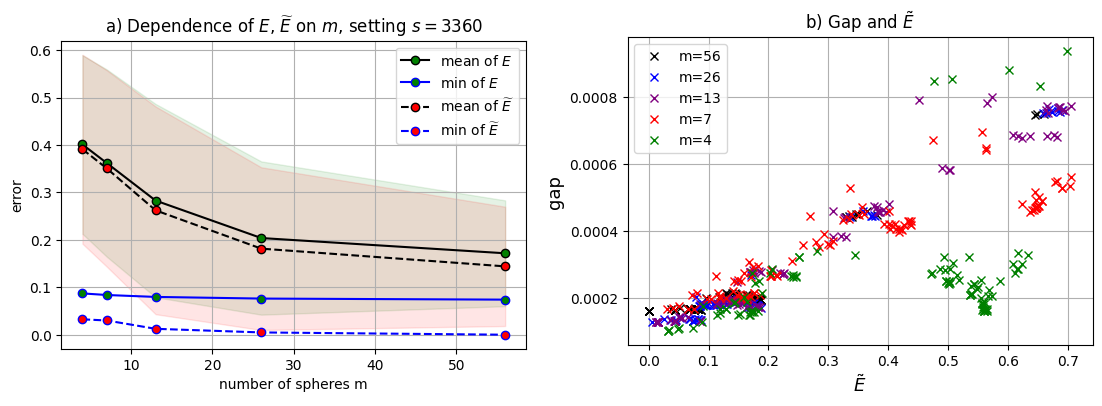}
    \caption{Correlation between a) number of spheres $m$ and errors $E$, $\tilde{E}$ and b) gap and error $\tilde{E}$ using a varying number ($m$) of spheres,
    setting $s=3360$, 100 globally random initialization for each $m$, for well-calibrated data.
    }
    \label{fig:m.error.and.gap}
\end{figure}
}\fi

\begin{figure}
  \begin{subfigure}{.45\textwidth}
      \centering
      \includegraphics[width=1.\linewidth]{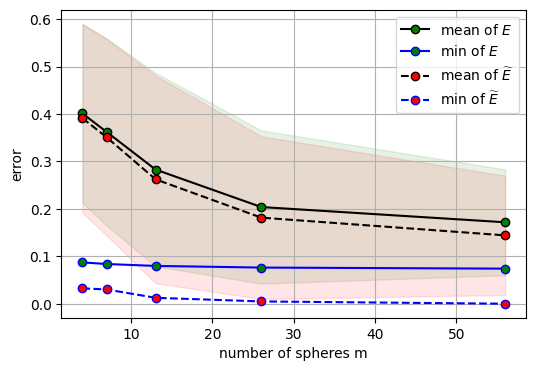}
      \caption{}
    \end{subfigure}
    \begin{subfigure}{.45\textwidth}
      \centering
      \includegraphics[width=1.\linewidth]{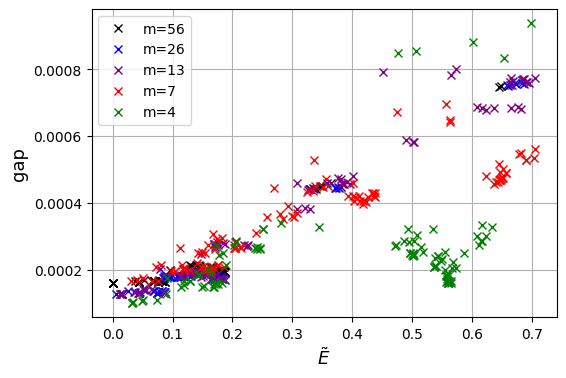}
      \caption{}
    \end{subfigure}
    \caption{Correlation between a) number of spheres $m$ and errors $E$, $\tilde{E}$ and b) gap and error $\tilde{E}$ using a varying number ($m$) of spheres,
    setting $s=3360$, 100 globally random initialization for each $m$, for well-calibrated data.
    }
    \label{fig:m.error.and.gap}
\end{figure}

    \subsubsection{Calibration error} \label{item:calibrated.error.investigate}
    In experiment, measuring 3D-POT data requires to tune the photon energy to a range of values and to measure the photoemission rate normalized to the amplitude of the incident light field. Consequently, a precise calibration of the - often fluctuating - extreme ultraviolet (EUV) light intensity if necessary. If fluctuations in the EUV power are not properly accounted for, \emph{calibration errors} may arise which scale the measured intensity on individually measured spheres with respect to the true momentum distribution of the orbital. Furthermore, it is well known that the plane-wave model of photoemission (Eq.~\ref{eq:plane_wave_model}) does not always provide a good prediction of the photon energy dependence of the photoelectron spectrum \cite{kern_simple_2023, gozem_photoelectron_2015}, which can lead to similar errors in the measured momentum distribution.
    To assess how sensitive the algorithm is to such model miss-specification, we now add EUV intensity calibration errors to the data.

    Let $m\in\{4,7,13,26,56\}$, the number of spheres.
    Let $\{L_\rho\}_{\rho=1}^m$ be a uniformly distributed random sample of lengths $m$ in the interval $(0,1)$, and let $\alpha\in (0,1)$.
    For $\rho=1,\dots,m$, the ideal data is distorted by following formula:
    \begin{equation}
        b'_{(i,j,l)}: = (1+\bigtriangleup a_\rho)b_{(i,j,l)}, \  \ (i,j,l)\in \mathbb{S}_\rho,
    \end{equation}
    where $\bigtriangleup a_\rho := \alpha(2L_\rho-1)$. Then $-\alpha\leq\bigtriangleup a_\rho\leq \alpha$ and we get
    \begin{equation}
      (1-\alpha)b_{(i,j,l)}\leq b'_{(i,j,l)} \leq (1+\alpha)b_{(i,j,l)}, \ \ (i,j,l)\in \mathbb{S}.
    \end{equation}
    We refer to $b'$ as calibrated data.
    Let $\alpha$ be chosen from the set $\{0.1, 0.2, 0.3, 0.6, 0.9\}$;  $\alpha$ has the interpretation of a measurement accuracy or miss-specification level, so these numbers correspond to \emph{maximal} noise or miss-specifications of 10\%, 20\%, 30\%, 60\%, and 90\% for each sphere, respectively. To clarify, for $\alpha=0.3$, the average photon flux at each photon energy is measured with an error margin of $\pm 30\%$.

    For each $m$, we generated $10$ samples $\{L_\rho\}_{\rho=1}^m$ yielding $50$ datasets for each value of $m$ across the respective levels of $\alpha$.
    We add $1$ ideal dataset (with $\alpha=0$) for each $m$.
    This results in a total of $255$ datasets.

    We study the behavior of the algorithm, for fixed algorithm parameters, under the realizations of calibration error as described above.

    The results displayed in Fig. \ref{fig:alpha.and.error} show the recovery performance, indicated by the error with the best numerical solution $\widetilde E$ defined by \eqref{eq:error.with.fixpoint}, over $25500$ reconstructions with a fixed $s=3360$ (i.e., $100$ globally random initializations per simulated dataset).
    The average number of iterations required for convergence for $m=4, 7, 13, 26, 56$ was about $2409, 485, 238, 155, 147$ iterations, respectively.

    \if{\begin{figure}
    \centering
    \includegraphics[width=\textwidth]{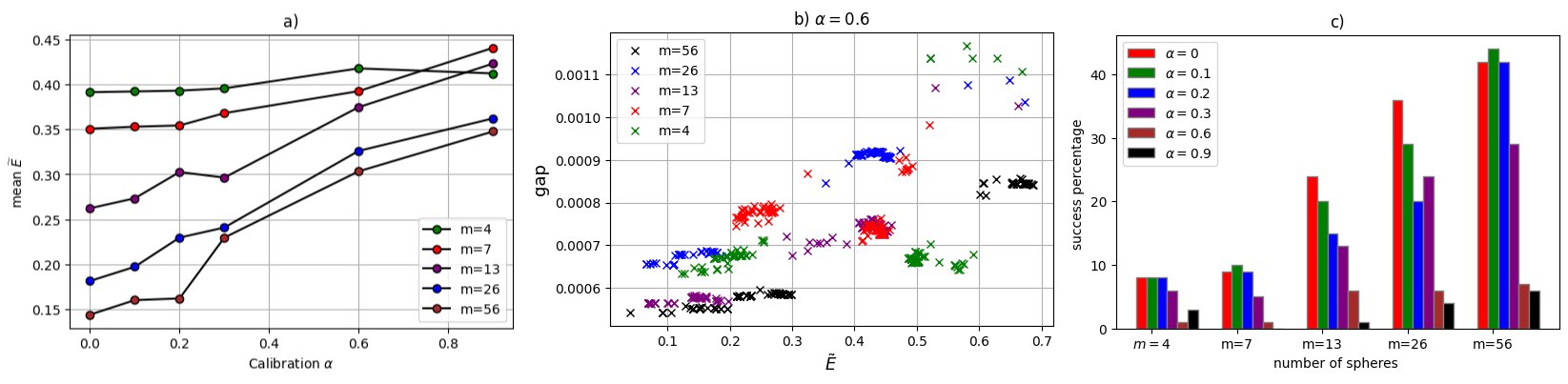}
    \caption{a) The average value of $\widetilde{E}$ as a function of calibration error $\alpha$; b) The correlation of gap and $\widetilde E$ with $\alpha=0.6$; c) The dependence of success percentage ($\widetilde E \leq 0.1$) on $m$ and $\alpha$.}
    \label{fig:alpha.and.error}
    \end{figure}
    }\fi
\begin{figure}
\begin{subfigure}{.45\linewidth}
\centering
\includegraphics[scale=0.5]{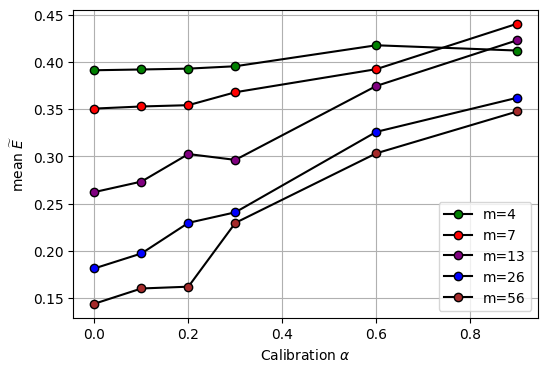}
\caption{}
\end{subfigure}%
\begin{subfigure}{.45\linewidth}
\centering
\includegraphics[scale=0.5]{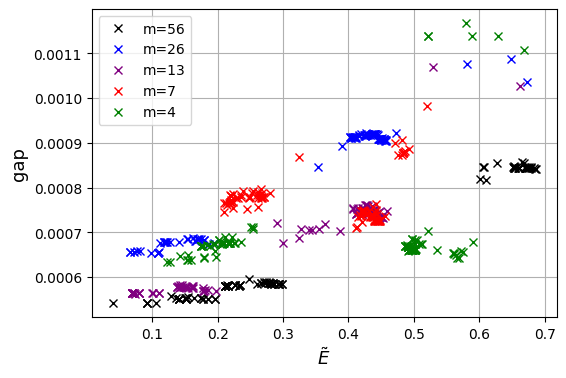}
\caption{}
\end{subfigure}\\[1ex]
\begin{subfigure}{\linewidth}
\centering
\includegraphics[scale=0.5]{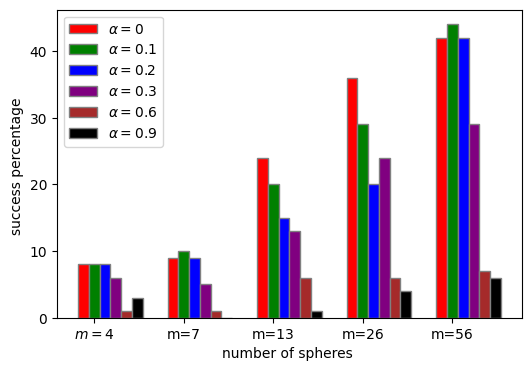}
\caption{}
\end{subfigure}
\caption{a) The average value of $\widetilde{E}$ as a function of calibration error $\alpha$; b) The correlation of gap and $\widetilde E$ with $\alpha=0.6$; c) The dependence of success percentage ($\widetilde E \leq 0.1$) on $m$ and $\alpha$.}
\label{fig:alpha.and.error}
\end{figure}

    Fig. \ref{fig:alpha.and.error}a) provides an overview of the average of $\widetilde{E}$ depending on $m$ and $\alpha$.
    As expected, we observe a decrease in mean error both as the data becomes more accurate (i.e. as $\alpha$ decreases) and as the data becomes richer (i.e. as $m$ increases).
    Note that for $m=4$, the dependence of the mean $\widetilde{E}$ on $\alpha$ is relatively weak. This surprising result has a relatively simple explanation: at $m=4$, the model already allows for so many false local minima, that the added complexity due to model miss-specification does not worsen the results significantly.

    However, these results also show that the mean error is not enough to characterize the algorithm performance. In the following, we define that a given reconstruction is successful if it has an error $\widetilde E \leq 0.1$ (i.e., when the error is below 10\%). At this error level, the reconstructed isosurfaces all match well to the true molecular orbital.
    Strikingly, in Fig. \ref{fig:alpha.and.error}c), we find very similar success rates for $\alpha \leq 0.3$ as for ideal data with $\alpha=0$. The success rates are slightly reduced, but a series of $100$ globally random initializations is still likely to retrieve several successful reconstructions, which can then be selected by means of the minimum gap.
    For larger values of $\alpha$, the data becomes significantly more inaccurate, resulting in a marked decrease in the percentage of success. Nevertheless, even in these cases a few successful reconstructions are found.
    From a numerical perspective, this shows that systematic model errors like calibration error do not eliminate physically meaningful reconstructions (interpreted as global minima), but such errors do introduce more local minima that are less meaningful.
    Furthermore, the miss-specification of the model affects the correlation between the gap and the error, as shown in Fig. \ref{fig:alpha.and.error}b. For larger miss-specification (here, $\alpha = 0.6$ with $m \leq 13$), the gap can no longer be used to select the best reconstructions, however we emphasize that this problem only arises for very large miss-specification level which are not realistic in experiment.
    We therefore conclude that the algorithm is robust in the presence of the calibration error. Optimal results of course require the best possible intensity calibration, but significant errors up to 30\% in the intensity calibration are not detrimental.

    \subsubsection{Number of electrons}\label{item:photon.error.investigate}
\begin{figure}
        \centering
        \includegraphics[width=\textwidth]{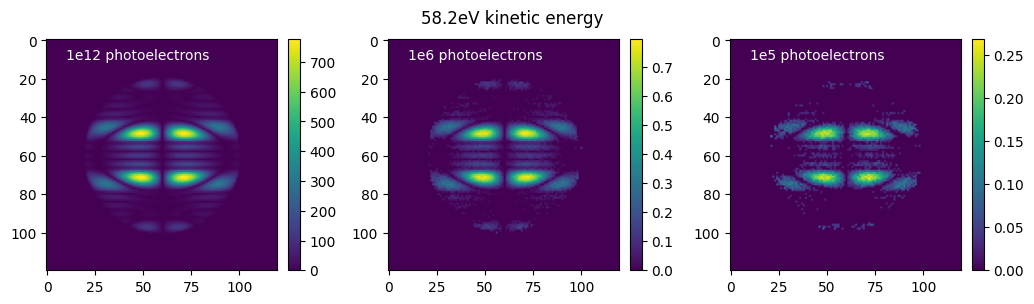}
        \caption{Exemplary simulated ARPES momentum maps from the $m=7$ data for different number of photoelectron counts.}
        \label{fig:ex.Poisson}
    \end{figure}

    Next to the calibration errors that were discussed above, an important source of model inconsistency is intensity noise in the measured momentum maps. Precisely how the momentum maps are acquired differs per experimental setup; however a common factor to all ARPES experiments is electron counting statistics. Here, we compare the effect of reducing signal-to-noise ratio of the data on the quality and reliability of the reconstructed orbitals by simulating datasets with varying total photoelectron counts. Other possible sources of experimental noise, such as background noise or the aforementioned calibration error, are neglected. Each of the datasets in the previous numerical experiments were generated with $10^{12}$ photoelectrons, which is large enough that the data can be assumed to be noise-free.  For this experiment, we now generate a dataset including $m=7$ photon energies using instead a total of $10^5$ detected photoelectrons (see Fig . \ref{fig:ex.Poisson} for an exemplary ARPES simulation).

    At this signal strength, the measurement data is significantly affected by Poissonian noise, and a significant fraction of the voxels that would otherwise indicate a low intensity now are zero: where in the noise-free case $6.4\%$ of the voxels in the data domain are non-zero, this reduces to $1.6\%$ at $10^5$ counts. Nevertheless, the orbital reconstruction is not significantly impacted: As before, we fixed $s=3360$, conducted reconstructions with $100$ globally random starting points, and found that in this case a success rate $7\%$ (i.e. with $\%(\widetilde E \leq 0.1)$).
    With $10^6$ incident electrons, where $3.1\%$ of the voxels is nonzero, the percentage success increases slightly to $8\%$.
    We emphasize that these total photoelectron counts are very easily achievable with all of the currently used photoemission orbital tomography detectors. Our results therefore indicate that efforts towards achieving higher data quality should be aimed at reducing other experimental noise factors, such as the determination and subtraction of background signals.

\section{Conclusions}
\label{sec:conclusions}
In this paper, we have addressed the challenge of reconstructing full 3D molecular orbitals from sparsely sampled 3D photoemission orbital tomography data.
We propose a feasibility model and utilize the cyclic projections method to reconstruct the orbital. The model incorporates prior constraints such as low-pass filtering, support, voxel sparsity, and symmetry properties.  Although this leads to an inconsistent feasibility model, we show that the global minimum closely approximates the true solution, which can be found with reasonably high probability by initializing the algorithm at several randomly chosen points and selecting the reconstruction with the lowest gap-value.

Remarkably, we find that the iterative orbital reconstruction method can deal with very sparse data and is robust in the presence of crucial experimental noise factors such as the light intensity calibration and the total signal strength.
In particular, a 3D-POT dataset covering only four photon energies in the EUV range already allows for accurate orbital reconstruction with over $8\%$ of the reconstructions achieving an error $\widetilde{E}$ within 90\% of the best possible value. This indicates that the orbitals to be reconstructed are sparse in {\em some } representation (e.g. spherical harmonics). Future research will investigate how to exploit such sparse representations.

Our results are especially promising in the context of laboratory-based, table-top photoemission orbital tomography, where high-harmonic generation EUV light sources provide access to only a limited number of photon energies. The reduced measurement requirements also provide other opportunities, including potentially the extension to \emph{time-resolved} 3D-POT \cite{Wallauer20sci, Neef23nat, Bennecke23arxiv, baumgartner_ultrafast_2022}.
A limitation of the method remains the relatively large variation in the results, as indicated by the standard deviation of errors in Fig.\ref{fig:s.m.and.error}.  This reflects a rather large variety of fixed points of the algorithm.  Future work will investigate the performance of the this algorithm on experimental data, as well as exploring alternative numerical methods for solving the feasibility problem whose fixed point sets are verifiably smaller.


\section*{Acknowledgments}
All authors were supported by the Deutsche Forschungsgemeinschaft (DFG, German Research Foundation) – Project-ID 432680300 – SFB 1456 Collaborative Research Center 1456 program - Mathematics of Experiment, project B01.


\if{\appendix
This subsection describes our approach to setting locally random initializations.
Let $x$ be any starting point in $\mathbb{C}^N$.
Let $\mathbb B(\widetilde \psi^*,r)$ be a small neighborhood of $\widetilde \psi^*$ with radius $r>0$.
Let $a\in(0,1)$. We have
$\|a\frac{x}{\|x\|}\|<1$
and $\|ra\frac{x}{\|x\|}\|<r.$
This implies
$$
    \widetilde\psi^*+ar\frac{x}{\|x\|}\in\mathbb B(\widetilde\psi^*,r), \forall x \in \mathbb{C}^N.
$$
Thus, for any random starting point $x\in\mathbb{C}^N$, by setting
\begin{equation}
    x^\prime = \widetilde\psi^*+ar\frac{x}{\|x\|},
\end{equation}
as a localization, we can map $x$ in the entire space to $x^\prime$ in the local neighborhood  of $\widetilde\psi^*$.
\if{
Fig. \ref{fig:stp.mapping} illustrated this in $\mathbb{R}^2$.

\begin{figure}
    \centering
    \includegraphics[scale=0.5]{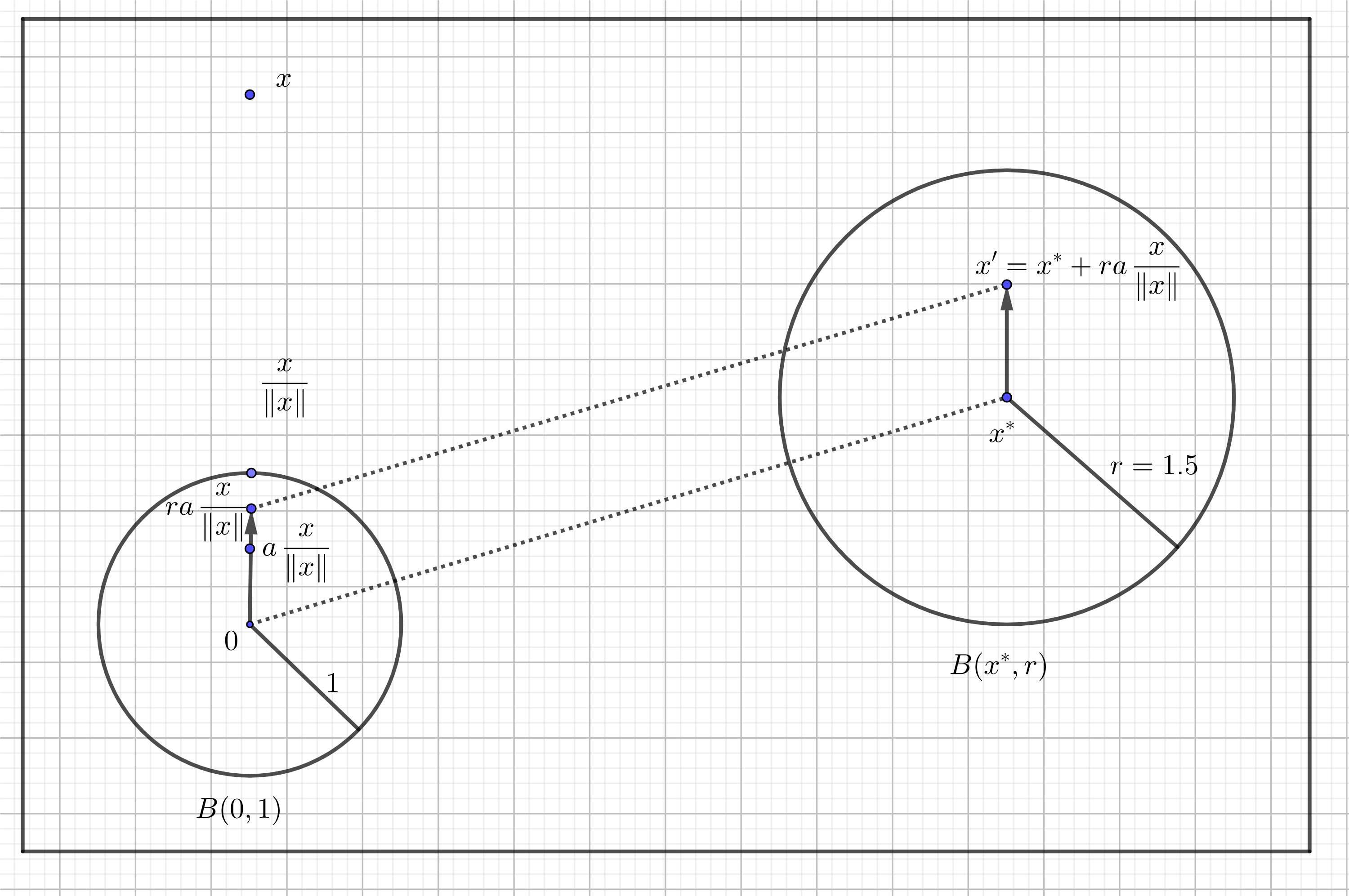}
    \caption{Illustration of mapping a globally random point to a neighborhood of a point in $\mathbb{R}^2$. Denote $x^*\equiv\widetilde\psi^*$. Setting $a=0.5, r=1.5$.}
    \label{fig:stp.mapping}
\end{figure}

\subsection*{Relationship between $E$ and $\widetilde E$}

    Similarly as in Fig. \ref{fig:Gap.Etilde.E500trials} we can use this dataset to verify the relationship between $E$ and $\widetilde{E}$; this is shown in Fig. \ref{fig:Etilde.and.E}).
    The yellow line represents the ideal case where $E = \widetilde{E}$; in this case the best feasible solution $\widetilde \psi^*$ would coincide with the true solution.
    While such linear behavior can be seen for $E \gtrsim 0.2$, a deviation from the ideal case appears at smaller values of $E$. This is a direct consequence of the inconsistent nature of the model \eqref{eq:intersect_form}. To reduce this cutoff therefore requires to reduce the discrepancy between $\psi^*$ and $\widetilde \psi^*$ by improving (or suitably relaxing) the constraints SYM, SR, SUPP, and LF. This falls beyond the scope of the current article, however.
\begin{figure}
    \centering
    \includegraphics[scale=0.7]{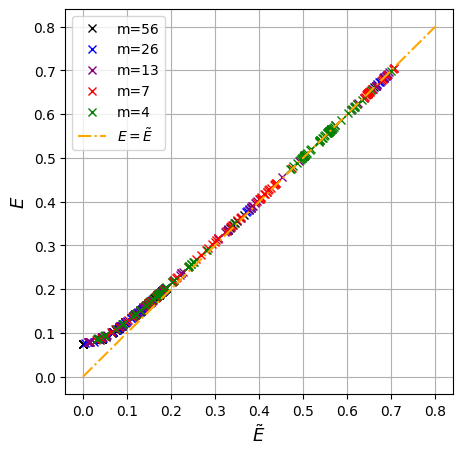}
    \caption{Caption}
    \label{fig:Etilde.and.E}
\end{figure}
}\fi
}\fi


\begin{thebibliography}{10}
\expandafter\ifx\csname url\endcsname\relax
  \def\url#1{{\tt #1}}\fi
\expandafter\ifx\csname urlprefix\endcsname\relax\def\urlprefix{URL }\fi
\providecommand{\eprint}[2][]{\url{#2}}

\bibitem{withers_x-ray_2021}
Withers P~J, Bouman C, Carmignato S, Cnudde V, Grimaldi D, Hagen C~K, Maire E,
  Manley M, Du~Plessis A and Stock S~R 2021 {\em Nature Reviews Methods
  Primers\/} {\bf 1} 1--21 ISSN 2662-8449 number: 1 Publisher: Nature
  Publishing Group
  \urlprefix\url{https://www.nature.com/articles/s43586-021-00015-4}

\bibitem{piccolomini_reconstruction_2018}
Piccolomini E~L, Coli V~L, Morotti E and Zanni L 2018 {\em Computational
  Optimization and Applications\/} {\bf 71} 171--191 ISSN 1573-2894
  \urlprefix\url{https://doi.org/10.1007/s10589-017-9961-2}

\bibitem{lustig_sparse_2007}
Lustig M, Donoho D and Pauly J~M 2007 {\em Magnetic Resonance in Medicine\/}
  {\bf 58} 1182--1195 ISSN 1522-2594 \_eprint:
  https://onlinelibrary.wiley.com/doi/pdf/10.1002/mrm.21391
  \urlprefix\url{https://onlinelibrary.wiley.com/doi/abs/10.1002/mrm.21391}

\bibitem{miao_extending_1999}
Miao J, Charalambous P, Kirz J and Sayre D 1999 {\em Nature\/} {\bf 400}
  342--344 ISSN 1476-4687 number: 6742 Publisher: Nature Publishing Group
  \urlprefix\url{https://www.nature.com/articles/22498}

\bibitem{Marchesini_x-ray_2003}
Marchesini S, He H, Chapman H~N, Hau-Riege S~P, Noy A, Howells M~R, Weierstall
  U and Spence J~C~H 2003 {\em Phys. Rev. B\/} {\bf 68}(14) 140101
  \urlprefix\url{https://link.aps.org/doi/10.1103/PhysRevB.68.140101}

\bibitem{puschnig_reconstruction_2009}
Puschnig P, Berkebile S, Fleming A~J, Koller G, Emtsev K, Seyller T, Riley J~D,
  Ambrosch-Draxl C, Netzer F~P and Ramsey M~G 2009 {\em Science\/} {\bf 326}
  702--706 ISSN 0036-8075, 1095-9203
  \urlprefix\url{https://science.sciencemag.org/content/326/5953/702}

\bibitem{marchesini_invited_2007}
Marchesini S 2007 {\em Review of Scientific Instruments\/} {\bf 78} 011301 ISSN
  0034-6748

\bibitem{jansen2020efficient}
Jansen G~S~M, Keunecke M, D{\"u}vel M, M{\"o}ller C, Schmitt D, Bennecke W,
  Kappert F, Steil D, Luke D~R, Steil S and Mathias S 2020 {\em New Journal of
  Physics\/} {\bf 22} 063012

\bibitem{candes_introduction_2008}
Candes E~J and Wakin M~B 2008 {\em IEEE Signal Processing Magazine\/} {\bf 25}
  21--30

\bibitem{dauth_orbital_2011}
Dauth M, Körzdörfer T, Kümmel S, Ziroff J, Wiessner M, Schöll A, Reinert F,
  Arita M and Shimada K 2011 {\em Phys. Rev. Lett.\/} {\bf 107} 193002
  \urlprefix\url{https://link.aps.org/doi/10.1103/PhysRevLett.107.193002}

\bibitem{zamborlini_multi-orbital_2017}
Zamborlini G, Lüftner D, Feng Z, Kollmann B, Puschnig P, Dri C, Panighel M,
  Di~Santo G, Goldoni A, Comelli G, Jugovac M, Feyer V and Schneider C~M 2017
  {\em Nat Commun\/} {\bf 8} 335 ISSN 2041-1723
  \urlprefix\url{https://www.nature.com/articles/s41467-017-00402-0}

\bibitem{yang_identifying_2019}
Yang X, Egger L, Hurdax P, Kaser H, Lüftner D, Bocquet F~C, Koller G, Gottwald
  A, Tegeder P, Richter M, Ramsey M~G, Puschnig P, Soubatch S and Tautz F~S
  2019 {\em Nat Commun\/} {\bf 10} 3189 ISSN 2041-1723
  \urlprefix\url{https://www.nature.com/articles/s41467-019-11133-9}

\bibitem{Repp_STM_2005}
Repp J, Meyer G, Stojkovi\ifmmode~\acute{c}\else \'{c}\fi{} S~M, Gourdon A and
  Joachim C 2005 {\em Phys. Rev. Lett.\/} {\bf 94}(2) 026803
  \urlprefix\url{https://link.aps.org/doi/10.1103/PhysRevLett.94.026803}

\bibitem{Soe_STM_2009}
Soe W~H, Manzano C, De~Sarkar A, Chandrasekhar N and Joachim C 2009 {\em Phys.
  Rev. Lett.\/} {\bf 102}(17) 176102
  \urlprefix\url{https://link.aps.org/doi/10.1103/PhysRevLett.102.176102}

\bibitem{itatani_tomographic_2004}
Itatani J, Levesque J, Zeidler D, Niikura H, Pépin H, Kieffer J~C, Corkum P~B
  and Villeneuve D~M 2004 {\em Nature\/} {\bf 432} 867--871 ISSN 1476-4687
  number: 7019 Publisher: Nature Publishing Group
  \urlprefix\url{https://www.nature.com/articles/nature03183}

\bibitem{vozzi_generalized_2011}
Vozzi C, Negro M, Calegari F, Sansone G, Nisoli M, De~Silvestri S and Stagira S
  2011 {\em Nature Phys\/} {\bf 7} 822--826 ISSN 1745-2481 number: 10
  Publisher: Nature Publishing Group
  \urlprefix\url{https://www.nature.com/articles/nphys2029}

\bibitem{graus_three-dimensional_2019}
Graus M, Metzger C, Grimm M, Nigge P, Feyer V, Schöll A and Reinert F 2019
  {\em Eur. Phys. J. B\/} {\bf 92} 80 ISSN 1434-6036
  \urlprefix\url{https://doi.org/10.1140/epjb/e2019-100015-x}

\bibitem{weis_exploring_2015}
Weiß S, Lüftner D, Ules T, Reinisch E~M, Kaser H, Gottwald A, Richter M,
  Soubatch S, Koller G, Ramsey M~G, Tautz F~S and Puschnig P 2015 {\em Nature
  Communications\/} {\bf 6} 8287 ISSN 2041-1723
  \urlprefix\url{https://www.nature.com/articles/ncomms9287}

\bibitem{Wallauer20sci}
Wallauer R, Raths M, Stallberg K, Münster L, Brandstetter D, Yang X, Güdde J,
  Puschnig P, Soubatch S, Kumpf C, Bocquet F~C, Tautz F~S and Höfer U 2021
  {\em Science\/} {\bf 371} 1056--1059

\bibitem{Neef23nat}
Neef A, Beaulieu S, Hammer S, Dong S, Maklar J, Pincelli T, Xian R~P, Wolf M,
  Rettig L, Pflaum J and Ernstorfer R 2023 {\em Nature\/} {\bf 616} 275--279
  ISSN 1476-4687 \urlprefix\url{https://doi.org/10.1038/s41586-023-05814-1}

\bibitem{baumgartner_ultrafast_2022}
Baumgärtner K, Reuner M, Metzger C, Kutnyakhov D, Heber M, Pressacco F, Min
  C~H, Peixoto T~R~F, Reiser M, Kim C, Lu W, Shayduk R, Izquierdo M, Brenner G,
  Roth F, Schöll A, Molodtsov S, Wurth W, Reinert F, Madsen A, Popova-Gorelova
  D and Scholz M 2022 {\em Nat Commun\/} {\bf 13} 2741 ISSN 2041-1723 number: 1
  Publisher: Nature Publishing Group
  \urlprefix\url{https://www.nature.com/articles/s41467-022-30404-6}

\bibitem{Bennecke23arxiv}
{Bennecke} W, {Windischbacher} A, {Schmitt} D, {Bange} J~P, {Hemm} R, {Kern}
  C~S, {D`Avino} G, {Blase} X, {Steil} D, {Steil} S, {Aeschlimann} M,
  {Stadtmueller} B, {Reutzel} M, {Puschnig} P, {Matthijs Jansen} G~S and
  {Mathias} S 2023 {\em arXiv e-prints\/} arXiv:2303.13904 (\textit{Preprint}
  \eprint{2303.13904})

\bibitem{yang_momentum-selective_2022}
Yang X, Jugovac M, Zamborlini G, Feyer V, Koller G, Puschnig P, Soubatch S,
  Ramsey M~G and Tautz F~S 2022 {\em Nature Communications\/} {\bf 13} 5148
  ISSN 2041-1723 number: 1 Publisher: Nature Publishing Group
  \urlprefix\url{https://www.nature.com/articles/s41467-022-32643-z}

\bibitem{lihuang_organic2d_2018}
Li Huang Y, Jie Zheng Y, Song Z, Chi D, S Wee A~T and Ying Quek S 2018 {\em
  Chemical Society Reviews\/} {\bf 47} 3241--3264 publisher: Royal Society of
  Chemistry
  \urlprefix\url{https://pubs.rsc.org/en/content/articlelanding/2018/cs/c8cs00159f}

\bibitem{keunecke_time-resolved_2020}
Keunecke M, Möller C, Schmitt D, Nolte H, Jansen G~S~M, Reutzel M, Gutberlet
  M, Halasi G, Steil D, Steil S and Mathias S 2020 {\em Review of Scientific
  Instruments\/} {\bf 91} 063905 type: Journal Article
  \urlprefix\url{https://aip.scitation.org/doi/abs/10.1063/5.0006531}

\bibitem{kutnyakhov_time-_2020}
Kutnyakhov D, Xian R~P, Dendzik M, Heber M, Pressacco F, Agustsson S~Y,
  Wenthaus L, Meyer H, Gieschen S, Mercurio G, Benz A, Bühlman K, Däster S,
  Gort R, Curcio D, Volckaert K, Bianchi M, Sanders C, Miwa J~A, Ulstrup S,
  Oelsner A, Tusche C, Chen Y~J, Vasilyev D, Medjanik K, Brenner G,
  Dziarzhytski S, Redlin H, Manschwetus B, Dong S, Hauer J, Rettig L, Diekmann
  F, Rossnagel K, Demsar J, Elmers H~J, Hofmann P, Ernstorfer R, Schönhense G,
  Acremann Y and Wurth W 2020 {\em Review of Scientific Instruments\/} {\bf 91}
  013109 ISSN 0034-6748

\bibitem{heber_multispectral_2022}
Heber M, Wind N, Kutnyakhov D, Pressacco F and Rossnagel K 2022 Multispectral
  time-resolved energy-momentum microscopy using high-harmonic extreme
  ultraviolet radiation \urlprefix\url{http://arxiv.org/abs/2203.11121}

\bibitem{kern_simple_2023}
Kern C~S, Haags A, Egger L, Yang X, Kirschner H, Wolff S, Seyller T, Gottwald
  A, Richter M, De~Giovannini U, Rubio A, Ramsey M~G, Bocquet F~m~c~C, Soubatch
  S, Tautz F~S, Puschnig P and Moser S 2023 {\em Phys. Rev. Res.\/} {\bf 5}(3)
  033075
  \urlprefix\url{https://link.aps.org/doi/10.1103/PhysRevResearch.5.033075}

\bibitem{kirschner_quantitative_2024}
Kirschner H, Gottwald A, Soltwisch V, Richter M, Puschnig P and Moser S 2024
  {\em Phys. Rev. A\/} {\bf 109}(1) 012814
  \urlprefix\url{https://link.aps.org/doi/10.1103/PhysRevA.109.012814}

\bibitem{HesseLukeNeumann14}
Hesse R, Luke D~R and Neumann P 2014 {\em {IEEE Trans. Signal. Process.}\/}
  {\bf 62} 4868--4881

\bibitem{giewekemeyer_high-dynamic-range_2014}
Giewekemeyer K, Philipp H~T, Wilke R~N, Aquila A, Osterhoff M, Tate M~W, Shanks
  K~S, Zozulya A~V, Salditt T, Gruner S~M and Mancuso A~P 2014 {\em Journal of
  Synchrotron Radiation\/} {\bf 21} 1167--1174 ISSN 1600-5775 publisher:
  International Union of Crystallography
  \urlprefix\url{//scripts.iucr.org/cgi-bin/paper?mo5086}

\bibitem{schroer_coherent_2008}
Schroer C~G, Boye P, Feldkamp J~M, Patommel J, Schropp A, Schwab A, Stephan S,
  Burghammer M, Schöder S and Riekel C 2008 {\em Physical Review Letters\/}
  {\bf 101} 090801 ISSN 0031-9007, 1079-7114
  \urlprefix\url{https://link.aps.org/doi/10.1103/PhysRevLett.101.090801}

\bibitem{BROEKMAN20051001}
Broekman L, Tadich A, Huwald E, Riley J, Leckey R, Seyller T, Emtsev K and Ley
  L 2005 {\em Journal of Electron Spectroscopy and Related Phenomena\/} {\bf
  144-147} 1001--1004 ISSN 0368-2048 proceeding of the Fourteenth International
  Conference on Vacuum Ultraviolet Radiation Physics
  \urlprefix\url{https://www.sciencedirect.com/science/article/pii/S0368204805000605}

\bibitem{BRANDSTETTER2021107905}
Brandstetter D, Yang X, Lüftner D, Tautz F~S and Puschnig P 2021 {\em Computer
  Physics Communications\/} {\bf 263} 107905 ISSN 0010-4655
  \urlprefix\url{https://www.sciencedirect.com/science/article/pii/S0010465521000461}

\bibitem{medjanik_direct_2017}
Medjanik K, Fedchenko O, Chernov S, Kutnyakhov D, Ellguth M, Oelsner A,
  Schönhense B, Peixoto T~R~F, Lutz P, Min C~H, Reinert F, Däster S, Acremann
  Y, Viefhaus J, Wurth W, Elmers H~J and Schönhense G 2017 {\em Nature
  Mater\/} {\bf 16} 615--621 ISSN 1476-4660
  \urlprefix\url{https://www.nature.com/articles/nmat4875}

\bibitem{GROsource}
Dinh T~L, Jansen G~S~M, Luke D~R, Bennecke W and Mathias S 2024 A minimalist
  approach to 3d photoemission orbital tomography: data and codes
  \urlprefix\url{https://publications.goettingen-research-online.de/}

\bibitem{luke2002optical}
Luke D~R, Burke J~V and Lyon R~G 2002 {\em SIAM review\/} {\bf 44} 169--224

\bibitem{luke2019optimization}
Luke D~R, Sabach S and Teboulle M 2019 {\em SIAM Journal on Mathematics of Data
  Science\/} {\bf 1} 408--445

\bibitem{marchesini2008ab}
Marchesini S 2008 {\em arXiv preprint arXiv:0809.2006\/}

\bibitem{kliuiev_algorithms_2018}
Kliuiev P, Latychevskaia T, Zamborlini G, Jugovac M, Metzger C, Grimm M,
  Sch\"oll A, Osterwalder J, Hengsberger M and Castiglioni L 2018 {\em Phys.
  Rev. B\/} {\bf 98}(8) 085426
  \urlprefix\url{https://link.aps.org/doi/10.1103/PhysRevB.98.085426}

\bibitem{russell2018quantitative}
Luke D~R, Thao N~H and Tam M~K 2018 {\em Mathematics of Operations Research\/}
  {\bf 43} 1143--1176

\bibitem{ACL16}
Aspelmeier T, Charitha C and Luke D~R 2016 {\em SIAM J. Imaging Sci.\/} {\bf 9}
  842--868

\bibitem{Damascelli_2004}
Damascelli A 2004 {\em Physica Scripta\/} {\bf 2004} 61
  \urlprefix\url{https://dx.doi.org/10.1238/Physica.Topical.109a00061}

\bibitem{gozem_photoelectron_2015}
Gozem S, Gunina A~O, Ichino T, Osborn D~L, Stanton J~F and Krylov A~I 2015 {\em
  The Journal of Physical Chemistry Letters\/} {\bf 6} 4532--4540 publisher:
  American Chemical Society
  \urlprefix\url{https://doi.org/10.1021/acs.jpclett.5b01891}

\end{thebibliography}
\providecommand{\newblock}{}

\end{document}